\documentclass{aa}  
\usepackage{graphicx}
\usepackage{txfonts}
\usepackage{lipsum}
\usepackage{lscape}
\usepackage{xcolor}
\usepackage{hyperref}
\usepackage[normalem]{ulem}
\usepackage{multirow}

\newcommand{\myfootref}[1]{\hyperlink{footnote.\getrefnumber{#1}}{\footnotemark[\getrefnumber{#1}]}}

\begin{document}

\title{A survey of ultra-compact high-state AM CVn binaries with ZTF and Gaia: New discoveries and observational constraints on Galactic space density}

\titlerunning{A survey of ultra-compact high-state AM CVn binaries with ZTF and Gaia}

\author{
Ilkham Galiullin\inst{1}
\and Antonio C. Rodriguez\inst{2,3}
\and Kareem El-Badry\inst{3}
\and Valery Suleimanov\inst{4}
\and Vladislav Dodon\inst{1}
\and Askar Sibgatullin\inst{1}
\and Warren R. Brown\inst{2}
\and Kevin Burdge\inst{5}
\and Jan van Roestel\inst{6,7}
\and Edo Berger\inst{2}
\and V. Ashley Villar\inst{2}
\and Ilaria Caiazzo\inst{7}
}
          
   \institute{Kazan Federal University, Kremlevskaya Str.18, 420008, Kazan, Russia\\
     \email{IlhIGaliullin@kpfu.ru}
        \and Center for Astrophysics $|$ Harvard \& Smithsonian, 60 Garden St, Cambridge, MA, 02138, USA
         \and Department of Astronomy, California Institute of Technology, 1200 E. California Blvd, Pasadena, CA, 91125, USA 
       \and Institut für Astronomie und Astrophysik, Kepler Center for Astro and Particle Physics, Universität Tübingen, Sand 1, 72076
        Tübingen, Germany
        \and Kavli Institute for Astrophysics and Space Research, Massachusetts Institute of Technology, Cambridge, MA 02139, USA
        \and Anton Pannekoek Institute for Astronomy, University of Amsterdam, 1090 GE Amsterdam, The Netherlands
        \and Institute of Science and Technology Austria, Am Campus 1, 3400 Klosterneuburg, Austria
             }      
    \authorrunning{Ilkham Galiullin et al.}
                 
   \date{Received XXX; accepted XXX}

  \abstract{ 
  Ultra-compact AM Canum Venaticorum (AM CVn)  binaries in the high-state of stable mass transfer ($\dot{M}\gtrsim 10^{-9}\,M_{\odot}\,\text{yr}^{-1}$) are expected to be among the loudest persistent sources for upcoming space-based gravitational wave (GW) observatories. These systems are typically characterized by orbital periods $P_{orb}\lesssim 30$ minutes. We present a systematic search for high-state AM CVn binaries in the Milky Way by targeting a region within the Gaia color-magnitude diagram, combined with Zwicky Transient Facility (ZTF) time-domain photometry. We focus on variable targets within that sample that had $5-30$ minute periods.  Our survey resulted in the discovery of three new high-state AM CVn binaries (ZTF~J1840$-$1742, ZTF~J2007$-$0527 and ZTF~J2111+3158) with respective orbital periods of 16.86, 18.63, and 16.71 minutes, and the recovery of three previously known systems. We obtain high-speed photometry and phase-resolved spectroscopy to confirm their nature. Their spectra show helium emission lines with an absence of detectable hydrogen. In two targets, the lines are double-peaked, and Doppler tomograms confirm the presence of an accretion disk. We estimate  mass accretion rates for these targets of the order of $10^{-9}\,M_{\odot }\,\text{yr}^{-1}$, and $3\sigma$ X-ray luminosity upper limits of the order of $\rm 10^{33}\ erg\ s^{-1}$. Based on this sample, we infer a local space density of high-state AM~CVn population in the Milky Way of  $\rho_0\sim 1.0 \times 10^{-8}~\text{pc}^{-3}$ to  $\rho_0\sim 2.7 \times 10^{-8}~\text{pc}^{-3}$ (for disk scale heights of $h_z=300,200$~pc, respectively), which represents only $2-5\%$ of the total Galactic AM CVn population density. Their birth rate, $R_{\text{birth}}\sim(2.4-4.3)\times 10^{-4}~\text{yr}^{-1}$ (for $h_z=300-200$~pc), is consistent with the total AM CVn birth rate, implying that most systems entering the high-state phase survive and evolve toward longer orbital periods. This local space density of high-state AM CVn binaries suggests that the millihertz GW observatories, Laser Interferometer Space Antenna (LISA) and TianQin, will detect about  $36\%$ and $19 \%$ of the total Galactic population, respectively, during their nominal 4-year missions with a signal-to-noise ratio $(S/N) \ge 5$. The Vera C. Rubin Observatory's Legacy Survey of Space and Time (LSST) is expected  to detect about $34-44\%$ of these systems over its 10-year survey at a median limit of $m_{r}\lesssim26.9$~mag.
  }

   \maketitle
   
\nolinenumbers

\section{Introduction}

AM Canum Venaticorum (AM~CVn) systems are ultra-compact binaries in which a white dwarf (WD) accretes matter from a Roche-lobe filling, semi-degenerate or degenerate helium-dominated donor. These systems typically have orbital periods in the $5-65$ minute range \citep[see for reviews,][]{2010PASP..122.1133S,2018A&A...620A.141R, 2025A&A...700A.107G}. Their optical spectra are characterized by helium emission lines and the absence of detectable hydrogen. Angular momentum loss in AM CVn binaries is dominated by gravitational wave (GW) emission, making them prominent sources for detection by future planned space-based GW observatories, such as Laser Interferometer Space Antenna (LISA) \citep{2017arXiv170200786A} and TianQin \citep{2016CQGra..33c5010L}. 

Three formation channels are proposed for AM CVn binaries: the WD channel, the helium star channel, and the evolved cataclysmic variable (CV) channel. 
In the WD channel \citep[e.g.,][]{1967AcA....17..287P,2007MNRAS.381..525D,2021ApJ...923..125W}, two common envelope episodes produce a double WD binary system consisting of a carbon-oxygen WD and a lower-mass helium WD. The lower-mass WD fills its Roche lobe at an orbital period of  $P_{orb}\sim 2-10 \, \text{minutes}$, at which point mass accretion begins. Shortly after the onset of accretion, the system reaches a peak mass accretion rate that can be high enough to trigger helium novae \citep{2015ApJ...805L...6S} or other thermonuclear events \citep[e.g.,][]{2023ApJ...951...28W}. If the mass transfer is stable, the system evolves toward longer orbital periods as an interacting AM~CVn binary. Unstable mass transfer leads to a merger, potentially resulting in the formation of R~CrB stars, hot subdwarfs, massive WDs, or Type~Ia supernovae \citep{1984ApJ...277..355W}.
The helium star donor channel follows a similar evolutionary path to the double WD channel, involving two common envelope (CE) phases that leave behind a carbon-oxygen WD with a low-mass, helium-burning donor star \citep[e.g.,][]{1987ApJ...313..727I, 2008AstL...34..620Y, 2023MNRAS.520.3187S}. Due to the donor's larger initial size, accretion begins at a longer period of $P_{orb}\sim 10-20 \, \text{minutes}$. Depending on the stability of the mass transfer, this channel can produce Type~Ia supernovae via a helium detonation that triggers a secondary carbon detonation \citep[e.g.,][]{2017ApJ...845...97B}, or it can result in stellar mergers.
In the evolved CV channel, the system initially evolves as a hydrogen-dominated CV after a single common envelope episode. The donor star in this system is significantly evolved before starting the mass transfer to the WD. After initiating the mass transfer, the outer hydrogen layer is stripped, revealing the helium-rich core. Evolved CVs pass the canonical period minimum for hydrogen-dominated CVs and continue to evolve towards a shorter orbital period \citep[e.g.,][]{1985SvAL...11...52T,2003MNRAS.340.1214P, 2021MNRAS.508.4106E, 2023MNRAS.520.3187S, 2023A&A...678A..34B}.

About 150 AM CVn systems are currently known, with a dozen identified as eclipsing systems \citep[e.g.,][for a review]{2025A&A...700A.107G}. Previous studies have obtained consistent results about the space density of AM CVn binaries in the Milky Way. Based on  the magnitude-limited sample of four systems from SDSS, \cite{2013MNRAS.429.2143C} estimated the space density of $\rho_0 =(5\pm3) \times 10^{-7}~\text{pc}^{-3}$. \cite{2018A&A...620A.141R} estimated a lower limit of $\rho_0 \ga 7 \times 10^{-8}~\text{pc}^{-3}$ using the Gaia distances of known objects identified across different surveys. Using the seven eclipsing systems from Zwicky Transient Facility (ZTF), \cite{2022MNRAS.512.5440V} obtained  $\rho_0 = (6^{+6}_{-2}) \times 10^{-7}~\text{pc}^{-3}$. \cite{2025PASP..137a4201R} used the sample of three systems within 150~pc, identified from SRG/eROSITA X-ray data, and estimated $\rho_0 = (5.5 \pm 3.7) \times 10^{-7}~\text{pc}^{-3}$. These values are an order of magnitude lower than the space density of hydrogen-dominated CVs in the Milky Way \citep[e.g., $\rho_0 = 4.8^{+0.6}_{-0.8} \times 10^{-6}~\text{pc}^{-3}$;][]{2020MNRAS.494.3799P}.

Depending on the orbital period and mass accretion rate, AM CVn binaries can exist in different states. At orbital periods of $P_{orb}\lesssim 30$ minutes and mass accretion rate of $\dot{M}\gtrsim 10^{-9}\,M_{\odot}\,\text{yr}^{-1}$, AM CVn binaries are generally classified as being in the high state. In high-state systems with orbital periods of $P_{orb}\sim 10-20$ minutes, the accretion disk remains in a persistently hot, ionized and thermally stable state. In contrast, systems with orbital periods of $P_{orb}\sim 20-40$ minutes are in the transition regime, typically exhibiting dwarf-nova-like outbursts driven by the thermal-viscous disk instability \citep{1983AcA....33..333S}. Some AM CVn binaries with orbital periods  $P_{orb}\lesssim 10$ minutes do not form an accretion disk, instead the accretion stream directly hits the WD surface \citep[direct-impact accretion; ][]{2004MNRAS.350..113M}. These systems have high X-ray luminosities, but they are rare due to their rapid evolutionary timescales. However, recent observations suggest, that even at orbital periods $P_{orb}\lesssim 10$ minutes, some AM CVn binaries can still host an accretion disk \citep{2024ApJ...977..262C}.  As AM CVn systems evolve toward longer orbital periods $P_{orb}\sim 40-50$ minutes, they transition into a low state. In that low state, the mass accretion rate drops significantly, and the accretion disk becomes cold with low intrinsic luminosity.

High-state AM CVn systems occupy the shortest-period end of the accreting WD population and are among the strongest Galactic GW sources, detectable with future missions such as LISA and TianQin \citep{2016CQGra..33c5010L,2018MNRAS.480..302K,2023LRR....26....2A}. Currently, about half of ``verification binaries'', binary systems that will be used to calibrate the sensitivity of millihertz GW observatories, are high-state AM CVns, with the other half being mostly detached double WD binaries. High mass transfer rate in high-state AM CVns enables efficient mass accumulation and helium burning on the WD, linking these systems to possible thermonuclear helium-triggered explosions \citep[Type ``.Ia'' SNe; e.g.,][]{2005ASPC..334..387S,2007ApJ...662L..95B}. Due to the intrinsic rarity of high-state AM CVn binaries compared to hydrogen-dominated CVs, efforts to empirically constrain their local space density remain challenging. While other surveys have already been used to successfully discover a sample of GW-emitting binaries \citep[primarily detached, \textit{non-accreting} double WDs; e.g.][]{2011ApJ...737L..23B,2019burdge,2020ApJ...905...32B}, the fainter nature of those objects complicates the recovery efficiency, therefore hindering efforts to constrain their space density. Since high-state AM CVns are more luminous due to ongoing accretion, their discovery at brighter magnitudes is possible with Gaia, enabling this survey.

In this paper, we present a systematic search for high-state AM CVn binaries based on a cross-match between the Gaia and ZTF data, resulting in the discovery of three new systems and the recovery of three previously known objects.  Using this sample, we provide the first empirical constraints on  the local space density of the high-state AM CVn population in the Milky Way and estimate the number of objects expected to be detectable with future observatories, such as LISA, TianQin and LSST. In Section \ref{sec:survey}, we present the search strategy and target selection. In Section \ref{sec:analysis}, we outline the spectroscopic follow-up observations of the new systems using Keck/LRIS and high-speed photometry with CHIMERA, and present the results of optical light curve, spectral, and spectral energy distributions (SEDs) analyses. In Section \ref{sec:discussion}, we discuss the properties of new objects, the local space density and the birth rate of high-state AM CVn population in the Milky Way, as well as the detectability of this population with LISA, TianQin, and LSST. We summarize our results in Section \ref{sec:summary}.

\begin{figure}[h]
    \centering
    \includegraphics[width=0.5\textwidth]{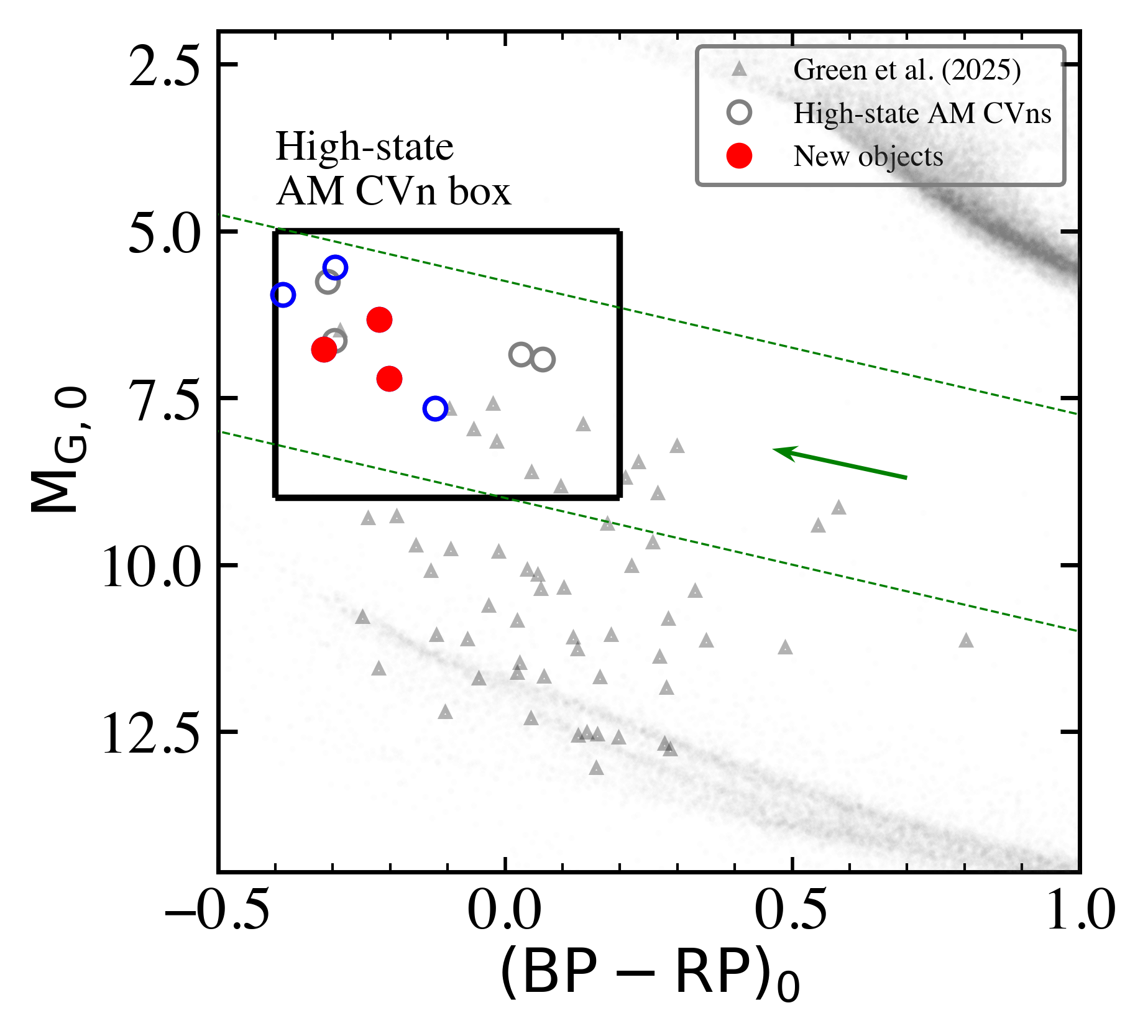}
    \caption{Extinction-corrected Hertzsprung--Russell (HR) diagram of AM~CVn binaries. Gaia objects within 100~pc are shown in gray. Gray triangles and circles represent all confirmed AM~CVns and high-state AM~CVns from the \citet{2025A&A...700A.107G} catalog, respectively, with parallaxes having signal-to-noise ratio $\varpi/\sigma_{\varpi} \ge 2$. Blue circles show known high-state AM~CVns recovered with our ZTF period search analysis. New systems from this study are shown as red filled circles. The black box shows the area used to select high-state AM~CVn candidates. Green dashed lines indicate the area used to query Gaia~DR3 data and  compile the initial sample. The green arrow represents the shift  of potential Gaia sources after extinction correction ($A_V=1$) and is included for illustrative purposes. }
    \label{fig:hr}
\end{figure}

\section{Survey motivation and target selection}
\label{sec:survey}

Our survey was motivated by Fig.~13 in \citet{2023rodriguez_amcvn}, where it appears that high-state AM~CVns (non-outbursting sources with orbital periods shorter than 30~minutes) cluster in a small region of the Hertzsprung--Russell (HR) diagram where few other objects are located. Fig.~\ref{fig:hr} shows the HR diagram for Gaia sources within 100~pc, along with confirmed AM~CVn binaries with parallaxes having signal-to-noise ratio $\varpi/\sigma_{\varpi} \ge 2$ from the \citet{2025A&A...700A.107G} catalog. Known high-state AM~CVns are located above the WD region and cluster at higher absolute magnitudes and bluer $(BP-RP)$ colors than long-period AM~CVns. We used this specific clustering to define our empirical selection criteria for high-state AM CVn candidates.

At the first step, we used the Gaia DR3 catalog \citep{2023A&A...674A...1G}. Distances were adopted from the geometric estimates of \cite{2021AJ....161..147B}. We selected sources with:

\begin{enumerate}
\setlength{\itemsep}{0.5em}
    \item $G<21$ mag; observed $G$-band magnitude is lower than 21 mag.
    \item $\tt RUWE < 1.4$; good fit to the astrometric observations.
    \item $\tt pmra / pmra\_error \geq 5$ and $\tt pmdec / pmdec\_error \geq 5$; proper motions in right ascension and declination directions are measured with a signal-to-noise ratio $\mu_{\alpha*}/\sigma_{\mu\alpha*} \ge 5$ and $\mu_{\delta}/\sigma_{\mu_\delta} \ge 5$.
    \item $\tt parallax\_over\_error \ge 2$; parallax is measured with a signal-to-noise ratio $\varpi/\sigma_{\varpi} \ge 2$.
    \item $\tt parallax > 0.1$; keeping only sources with distances $d<10$~kpc (or $\varpi > 0.1$~mas).
\end{enumerate}

We also included specific cuts on observed absolute magnitudes to include sources within the area of high-state AM CVns on the HR diagram with:

 \begin{enumerate}
 \setlength{\itemsep}{0.5em}
     \item $M_{G} \leq$   9 + $2\times (BP-RP)$.
     \item $M_{G} \geq$   5.75 + $2\times (BP-RP)$.
    \item $-0.5 \leq (BP-RP) \leq 1$.
 \end{enumerate}

These queries yielded $1\ 427\ 184$ sources, including the seven known high-state AM CVn binaries: ES Cet, SDSS J1351$-$0643, AM CVn, SDSS J1908+3940, HP Lib, CXOGBS J1751$-$2940, TIC 37889811 (see Appendix~\ref{app:list}). Fig.~\ref{fig:hr} (green dashed lines) shows the specific regions used to query our objects from Gaia~DR3. We used such an extended region to account for the shift of objects on the HR diagram due to extinction correction (see the illustrative green arrow in Fig.~\ref{fig:hr}).

Next, we kept only objects in the ZTF footprint with a declination of $\delta\geq -28^\circ$, which resulted in $564\ 869$ objects. For all targets, we estimated the extinction value $E(B-V)$ using a combination of three dust maps \citep[{\tt Combined19map};][]{2003A&A...409..205D,2006A&A...453..635M,2019ApJ...887...93G}, implemented in the {\tt mwdust}\footnote{\url{https://github.com/jobovy/mwdust}} package \citep{2016ApJ...818..130B}. We computed the extinction-corrected absolute $M_{G,0}$ magnitudes, $(BP-RP)_0$ colors and selected only sources in the empirical "high-state AM CVn box" (see Fig.~\ref{fig:hr}), defined as:
 \begin{enumerate}
 \setlength{\itemsep}{0.5em}
     \item $5 \leq M_{G,0} \leq 9$.
     \item $-0.4 \leq (BP-RP)_0 \leq 0.2$.
 \end{enumerate}
This resulted in $8546$ objects. We cross-matched each source with the ZTF archive, keeping only sources with $N > 50$ points in the light curve. This resulted in 5681 sources, including the five known high-state AM CVn binaries: ES Cet, SDSS J1351$-$0643, AM CVn, SDSS J1908+3940, and HP Lib. The declinations of CXOGBS J1751$-$2940 and TIC 37889811 ($\delta < -28^{\circ}$) are too far south for the ZTF survey footprint. We removed any potential outbursting sources showing a magnitude change of $\ge 2$~mag. We conducted the period search using a Lomb-Scargle periodogram in the {\tt gatspy}\footnote{\url{https://www.astroml.org/gatspy/}} package over the $5-30$ minutes range, and kept only sources with a period significance $> 30$ that passed visual inspection. The significance of the variability was defined as the ratio of the highest peak to  the median absolute deviation of the periodogram.

\begin{figure*}
    \centering
    \includegraphics[width=1\textwidth]{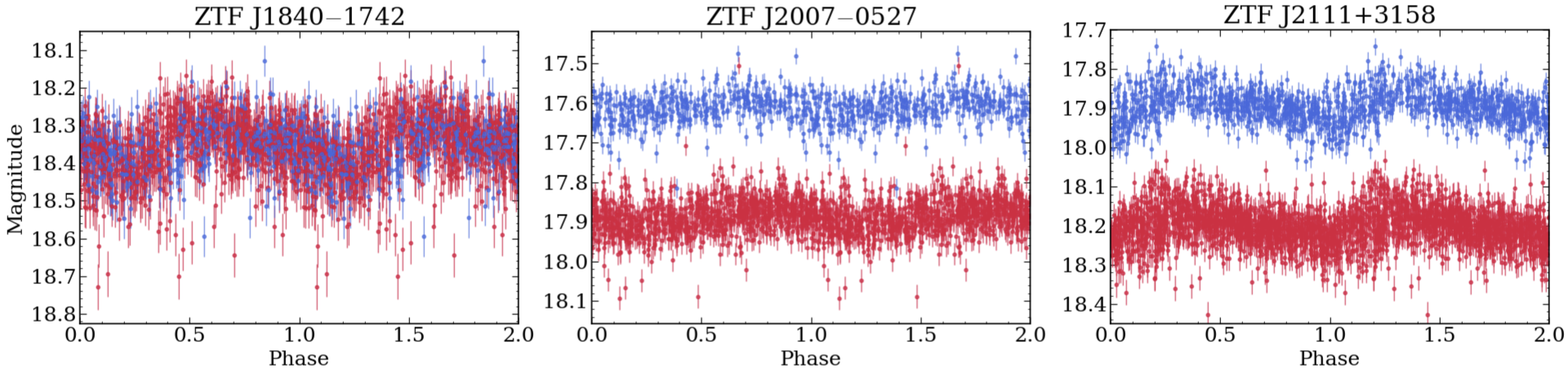}
    \caption{ The ZTF light curves for three new high-state AM~CVn binaries (ZTF J1840$-$1742, ZTF J2007$-$0527, and ZTF J2111+3158) in the $r$ and $g$ filters (red and blue colors), folded with their respective periods ($16.57$, $18.76$, and $17.02$ minutes).}
    \label{fig:lcs}
\end{figure*}

We found eight objects that passed all previous criteria, recovering three known variable high-state AM CVns from the \citet{2025A&A...700A.107G} catalog: ES Cet, SDSS J1351$-$0643, and HP Lib (see Appendix~\ref{app:list}). We discovered three new high-state AM CVn candidates: ZTF J1840$-$1742, ZTF J2007$-$0527, and ZTF J2111+3158, with periods of $16.57 \pm 0.25$, $18.76 \pm 0.25$, and $17.02 \pm 0.25$ minutes, respectively. The errors in the periods correspond to the time resolution of the ZTF data. Fig.~\ref{fig:lcs} shows the phase-folded ZTF light curves ($r$ and $g$ filters) for these three new AM CVn candidates. Two other objects discovered are not high-state AM CVns.

We note that the ”high-state AM~CVn box” was defined empirically based on the location of known high-state AM~CVns. We experimented by extending the size of this box by a factor of two to search for additional short-period variables. We found no additional short-period binaries with our ZTF period search analysis. Two known systems, AM CVn and SDSS J1908+3940, passed both our Gaia and ZTF criteria but showed no detectable variability in their ZTF light curves. We discuss these sources in Section~\ref{sec:population}.

\section{New Systems}
\label{sec:analysis}

\begin{figure*}
    \centering
    \includegraphics[width=0.85\textwidth]{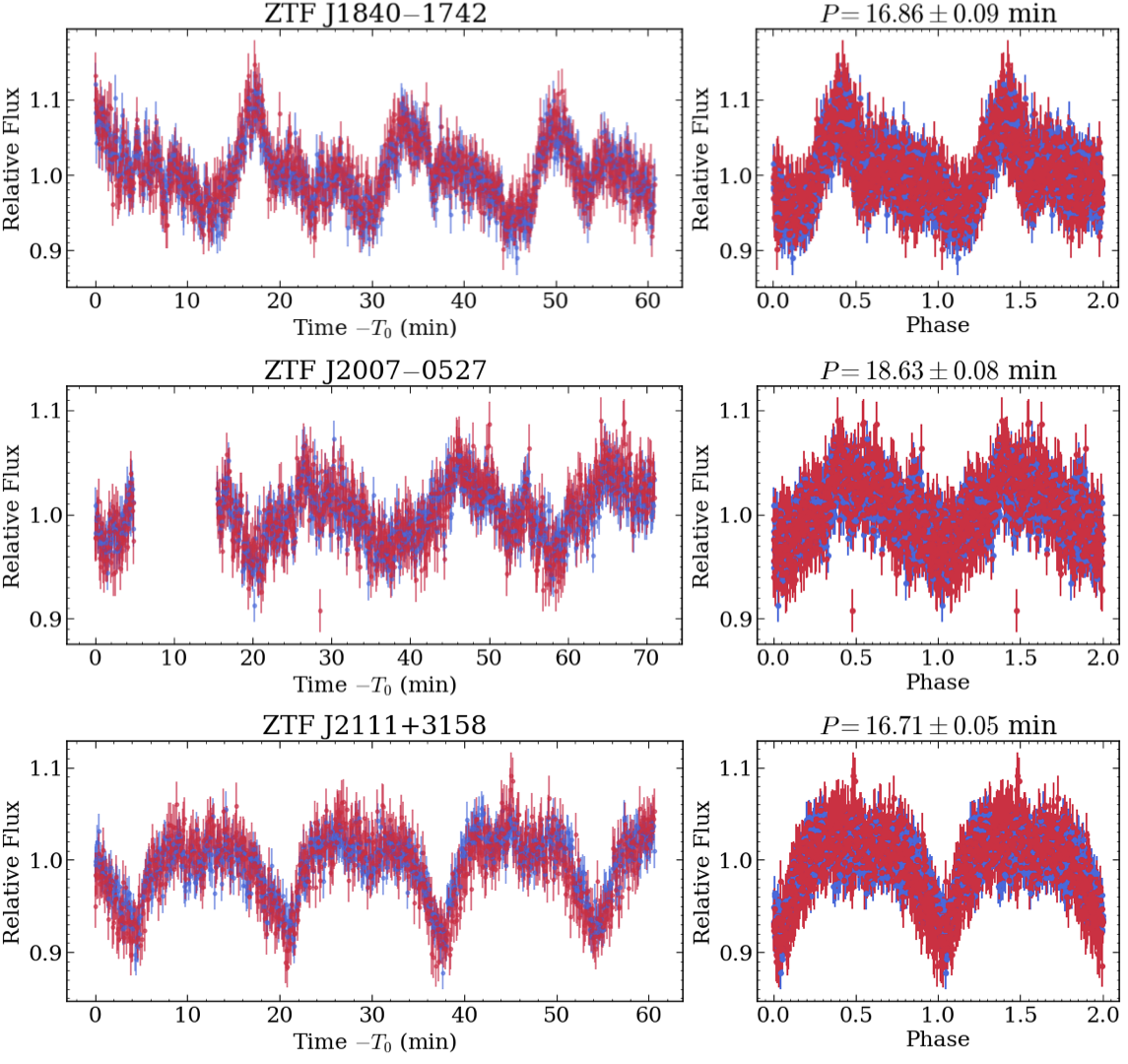}
    \caption{ CHIMERA high-speed photometry light curves for the three new high-state AM~CVns in the $r$ and $g$ filters. Left: Full light-curves spanning multiple orbital periods from the start of the observations. Right: Light-curves, folded using their refined orbital periods. }
    \label{fig:chimera}
\end{figure*}

\subsection{High-speed photometry and orbital periods}

We obtained high-speed photometry for  ZTF~J1840$-$1742, ZTF~J2007$-$0527, and ZTF~J2111+3158 on 29 August 2024, using the Caltech High-speed Multi-color Camera (CHIMERA; \citep{chimera}) on the Hale 200-inch telescope. In all cases, $r$- and $g$-filter data were simultaneously acquired at a 10-second cadence. We calibrated all CHIMERA data (bias-subtracted and flat-fielded) using standard techniques in the {\tt PyCHIMERA}\footnote{\url{https://github.com/caltech-chimera/PyChimera}} pipeline. The observations with CHIMERA covered approximately $60-70$~minutes, covering to $3-4$ orbital cycles (see Fig.~\ref{fig:chimera}, left panels). We again used the Lomb--Scargle periodogram to analyze the periodic signals in the CHIMERA data and to better constrain the periods identified in the ZTF data. We searched for periods in the combined $r$- and $g$-filter data within a range of $5-30$~minutes. To estimate period uncertainties, we used a Monte Carlo approach, resampling the flux within its errors assuming a Gaussian distribution. We performed 1000 iterations, determining the best-fit period for each iteration. The final orbital periods and their $1\sigma$ uncertainties were derived from the resulting distributions. We refined the periods of ZTF~J1840$-$1742, ZTF~J2007$-$0527, and ZTF~J2111+3158 to $16.86\pm0.09$, $18.63\pm0.08$, and $16.71\pm0.05$~minutes, respectively (see Fig.~\ref{fig:chimera}, right panels).

\subsection{Optical spectroscopy and Doppler tomography}

\begin{figure*}
    \centering
    \includegraphics[width=1\linewidth]{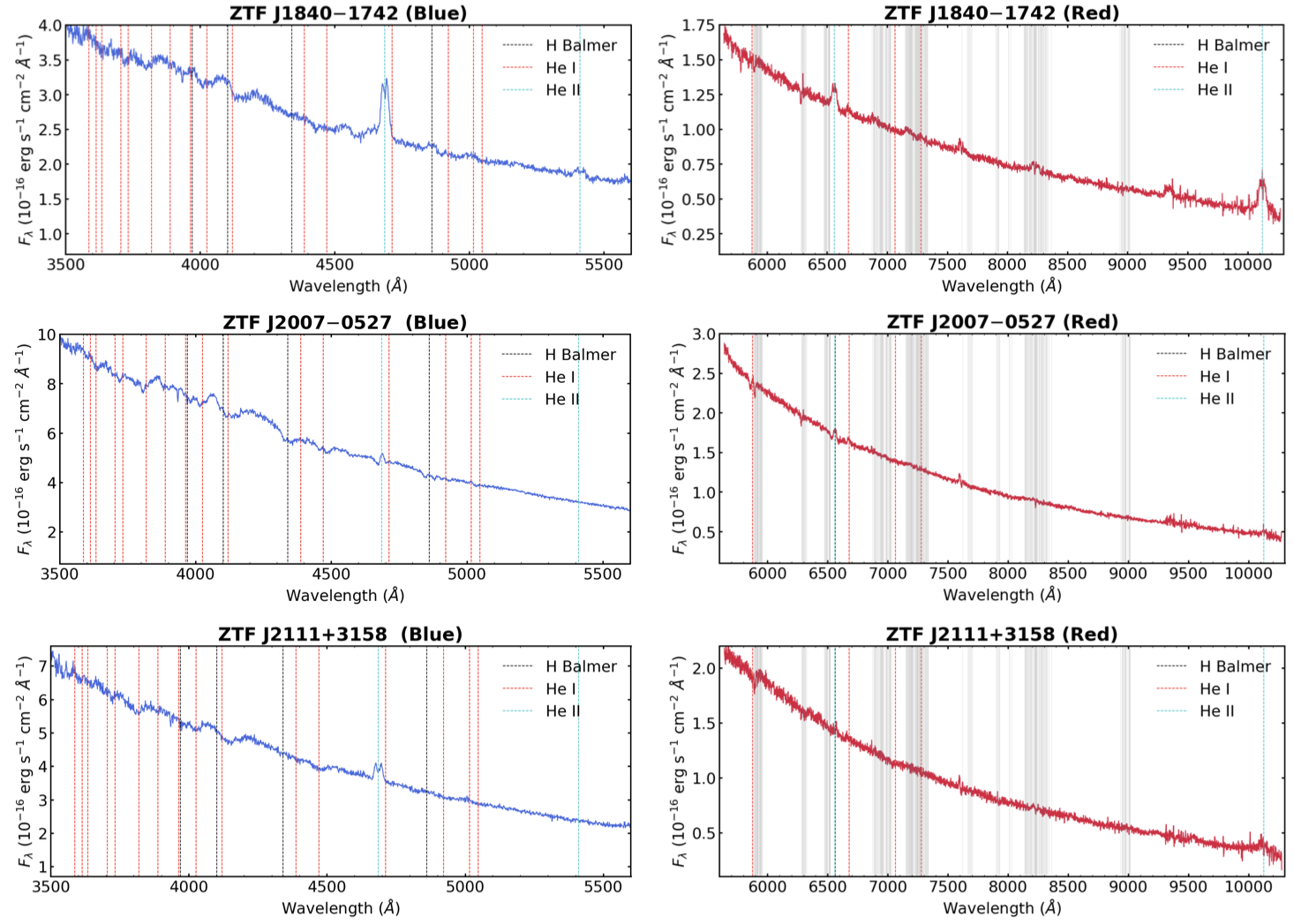}
    \caption{ Keck I/LRIS optical spectra of new high-state AM~CVns, separated by blue (left) and red (right) arms. The vertical lines show the locations of different emission lines. Gray shaded regions indicate the location of telluric lines. The HeII~$\lambda$4686~\AA\ emission lines of ZTF J1840$-$1742 and ZTF J2111+3158 are double-peaked. The weak emission lines near 6560~\AA\ (for  ZTF~J1840$-$1742 and ZTF~J2007$-$0527) correspond to the HeII (6560.1~\AA) emission line rather than H$\alpha$ (6562.8~\AA). }
    \label{fig:spectra}
\end{figure*}

\begin{figure*}
    \centering
    \includegraphics[width=\textwidth]{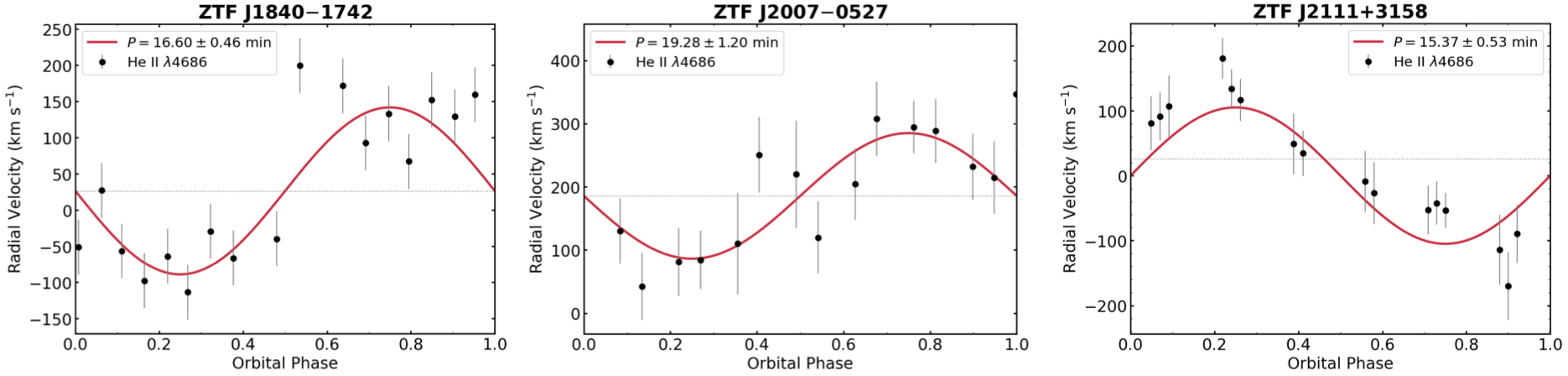}
    \caption{ RV measurements for the HeII~$\lambda$4686~\AA\ emission lines of our targets. The solid red line shows the best-fit RV model. The horizontal line shows the systemic velocity. The spectroscopic period determined from the RV analysis is consistent with the photometric period from the CHIMERA high-speed photometry data.}
    \label{fig:rvs}
\end{figure*}

\begin{figure*}
    \centering
    \includegraphics[width=1\textwidth]{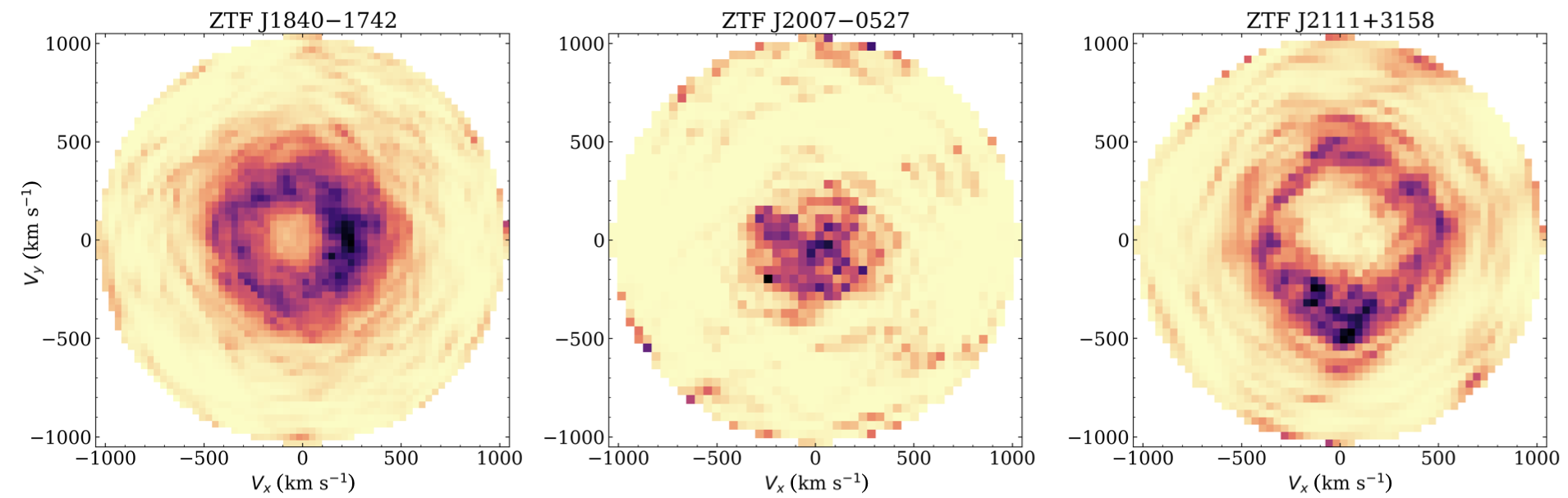}
    \caption{ Doppler tomograms of our targets, constructed for the HeII~$\lambda$4686~\AA\ emission lines. A disk structure is observed for ZTF~J1840$-$1742 and ZTF~J2111+3158. ZTF~J2007$-$0527 shows an extended feature resembling a centrally-filled disk. }
    \label{fig:doppler}
\end{figure*}

We performed the spectroscopic follow-up of AM CVn candidates with the Keck I telescope using the Low-Resolution Imaging Spectrometer (LRIS; \citep{lris}). We used the 600/4000 grism on the blue side with 2$\times$2 binning (spatial and spectral), and the 400/8500 grating on the red side with 2$\times$1 binning. We used a 1.0$\arcsec$ slit, and the seeing during the observations was 0.7$\arcsec$, resulting in minimal slit losses. Data calibration was performed using the {\tt lpipe} pipeline \citep{2019perley_lpipe}, optimized for LRIS imaging and long-slit spectroscopy, following standard procedures, including wavelength calibration with internal lamps, flat-field correction, and cosmic ray cleaning. Phase-resolved spectroscopy covering the entire orbit was obtained with LRIS on 9 July 2024 for ZTF J2111+3158 , and on 9 October 2024 for ZTF J1840$-$1742 and ZTF J2007$-$0527. Fig.~\ref{fig:spectra} shows the phase-averaged optical spectra for our targets, with the left and right panels corresponding to the blue and red arms, respectively. We do not observe any signatures of the donor stars, which indicates that they are colder than the accretion disk observed at optical wavelengths in each system.

To determine the spectroscopic orbital periods of our targets and compare them to the photometric periods from CHIMERA high-speed photometry, we performed a radial velocity (RV) analysis. We measured the RVs from the prominent HeII~$\lambda$4686~\AA\ emission line by fitting single or double Gaussian line profiles. We used {\tt scipy.optimize.curve\_fit} module to fit the line profiles and estimate the period uncertainties. We determined the spectroscopic periods of ZTF~J1840$-$1742, ZTF~J2007$-$0527, and ZTF~J2111+3158 to be $16.60\pm0.46$, $19.28\pm1.20$, and $15.37\pm0.53$~minutes, respectively (see Fig.~\ref{fig:rvs}). The large uncertainties in the spectroscopic periods and RV measurements were caused by the moderate signal-to-noise ratio of the HeII~$\lambda$4686~\AA\ emission line in each phase. The spectroscopic periods are consistent within their uncertainties with the photometric periods. We therefore interpret the more precisely constrained photometric periods as the orbital periods of the systems.

We observe weak emission lines near 6560~\AA\ for ZTF~J1840$-$1742 and ZTF~J2007$-$0527, which match either the H$\alpha$ (6562.8~\AA) line or the HeII Pickering series (6560.1~\AA). Similar features have been observed in several AM~CVn systems and interpreted as HeII (6560.1~\AA) emission line rather than H$\alpha$ line \citep[e.g.,][]{2018MNRAS.477.5646G, 2019MNRAS.485.1947G, 2021A&A...645A.114B}. We followed a similar approach to \citet{2018MNRAS.477.5646G} and measured the RV difference between these two possibilities. We first assumed that the observed 6560~\AA\ line originated from H$\alpha$ (6562.8~\AA) line and computed the RVs, then repeated the computation assuming the line originated from the HeII (6560.1~\AA) emission line. For ZTF~J1840$-$1742, we measured RV shifts of $-200 \pm 37$~km/s for the H$\alpha$ case and $-77 \pm 39$~km/s for the HeII case. For ZTF~J2007$-$0527, we obtained $-242 \pm 43$~km/s for H$\alpha$ and $-113 \pm 43$~km/s for HeII. Since the H$\alpha$ assumption yields a larger RV offset than the HeII case, it indicates the line originates from HeII (6560.1~\AA). The absence of other Balmer series lines (H$\beta$ and H$\gamma$) further supports that the observed 6560~\AA\ emission feature belongs to the HeII Pickering series rather than the H$\alpha$ line in the optical spectra of both ZTF~J1840$-$1742 and ZTF~J2007$-$0527.

The prominent HeII~$\lambda$4686~\AA\ emission lines are double-peaked for ZTF J1840$-$1742 and ZTF J2111+3158, indicating the presence of an accretion disk in these systems. We constructed Doppler tomograms for our targets using {\tt PyDoppler} \citep{2021ascl.soft06003H} for the HeII~$\lambda$4686~\AA\ emission lines. Fig.~\ref{fig:doppler} shows the Doppler tomograms for our targets. We observe a disk structure for ZTF J1840$-$1742 and ZTF J2111+3158.  For ZTF J2007$-$0527, the HeII~$\lambda$4686~\AA\  emission line is single-peaked, resulting in an extended feature resembling a centrally-filled disk in the Doppler tomogram. Our targets extend the sample of short-period AM~CVns harboring an accretion disk instead of undergoing direct impact accretion (see Table~\ref{tab:known}).

\begin{table}[ht]
\caption{Parameters of the new high-state AM~CVn systems.}
\label{tab:params}
\centering
\renewcommand{\arraystretch}{1.2} 
\begin{tabular*}{\columnwidth}{@{\extracolsep{\fill}} l | c c c}
\hline\hline
Object Name & J1840$-$1742 & J2007$-$0527 & J2111+3158 \\
\hline
R.A. (deg) & 280.2187 & 301.8123 & 317.8309 \\
Decl. (deg) & $-17.7136$ & $-5.4645$ & 31.9795 \\
$P_{orb}$ (min) & $16.86\pm0.09$ & $18.63\pm0.08$ & $16.71\pm0.05$ \\
$d$ (pc) & $1528^{+527}_{-302}$ & $1255^{+164}_{-158}$ & $1348^{+252}_{-167}$ \\
$E(B-V)$ & $0.33\pm 0.01$ & $0.15\pm 0.01$ & $0.04\pm 0.01$ \\
\hline
$R_{out}$ ($R_\odot$) & $0.17\pm0.05$ & $0.07\pm0.01$ & $0.09\pm0.02$ \\
$\log\dot{M}^a$ & $-8.91\pm0.29$ & $-8.64\pm0.12$ & $-9.20\pm0.17$ \\
$L_X$ ($\rm erg\ s^{-1}$)$^{b}$ & $\lesssim 2.1 \times 10^{33}$ & $\lesssim 5.2 \times 10^{33}$ & $\lesssim 2.0 \times 10^{33}$ \\
\hline
\end{tabular*}
\flushleft
Notes: $(a)$ -- in units of $M_\odot$ yr$^{-1}$; $(b)$ -- 3$\sigma$ upper limit for the 0.2--12 keV energy band.
\end{table}

\begin{figure*}
    \centering
    \includegraphics[width=1\textwidth]{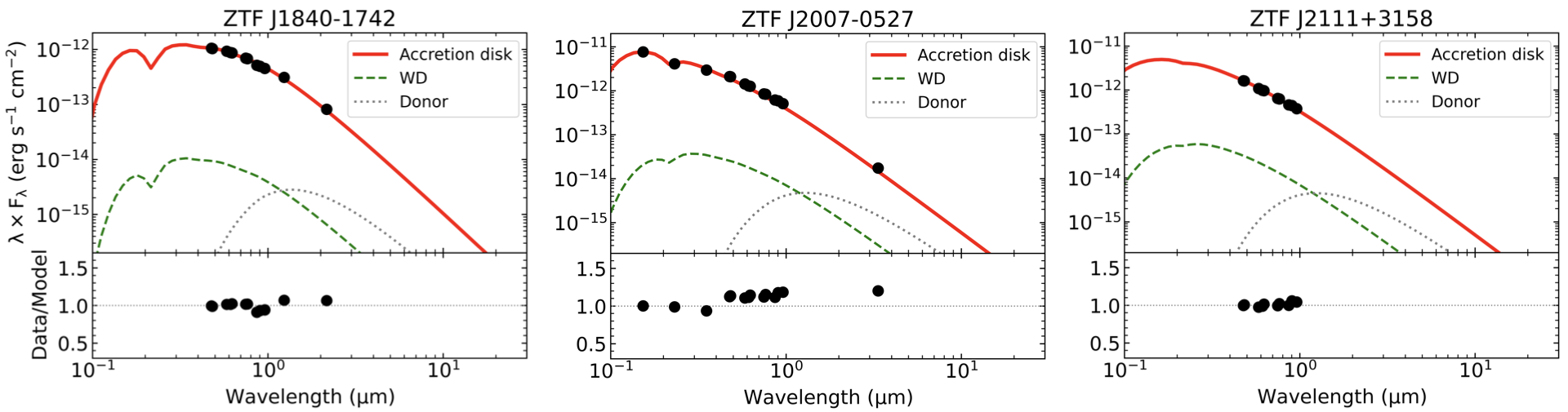}
    \caption{The observed SED of each target (top panels) and the ratio between the data and the model (bottom panels). The best-fit accretion disk models are shown in red. For illustrative purposes,  the possible contributions of the WD (green dashed lines; $T_{\rm eff} = 16000$~K, $R_{\rm WD} = 0.01~R_{\odot}$) and a cold donor (gray dotted lines; $T_{\rm eff} = 3000$~K, $R_{\rm donor} = 0.07~R_{\odot}$) are shown. Their combined contribution to the observed SED is negligible compared to the accretion disk component alone.
 }

    \label{fig:sed}
\end{figure*}

\subsection{SED modeling and mass accretion rate}

We constructed the observed spectral energy distributions (SEDs) of our objects using the VizieR photometry tool\footnote{\url{http://vizier.cds.unistra.fr/vizier/sed/}}. Photometric data were queried from available catalogs using a search radius of  $\rm 3 \arcsec$ around each target position: GALEX NUV, FUV \citep{2017ApJS..230...24B}; SDSS $u,g,r,i,z$ \citep{2015ApJS..219...12A}; Gaia $G$ \citep{2023A&A...674A...1G}; PAN-STARRS $g,r,i,z,y$ \citep{2020ApJS..251....6M}; 2MASS $J,H,K_s$ \citep{2006AJ....131.1163S}; and CatWISE $W_1$, $W_2$ \citep{2020ApJS..247...69E}. 

The SED of high-state AM CVn binaries is primarily dominated by emission from the accretion disk \citep[e.g.,][]{2018A&A...620A.141R}. %This allows us to estimate the accretion rate in the system. 
We assumed that the accretion disk is in a steady state, with its effective temperature varying with radius from the central WD according to the Shakura-Sunyaev accretion disk temperature profile \citep{1973A&A....24..337S}:
\begin{gather}
T(R) = \left[ \frac{3 G M_{WD} \dot{M}}{8 \pi \sigma R^3} \left( 1 - \sqrt{\frac{R_{\rm in}}{R}} \right) \right]^{1/4}, 
    \label{eq:tdisk}
\end{gather}
where the accretion disk model included the following parameters: the inner radius ($R_{in}$), the mass accretion rate ($\dot{M}$), the inclination angle of the system ($i$), and the WD mass ($M_{WD}$). 

We assumed that the accretion disk consists of concentric, circular annuli, each emitting as a blackbody source. We computed the total flux from the accretion disk by integrating the emission across all annuli, spanning from the inner to the outer disk radii ($R\in [R_{in};R_{out}]$):
\begin{equation}
    F_{\lambda} = \frac{\cos i}{d^2} \int_{R_{in}}^{R_{out}} B_{\lambda}(T(R)) \, 2\pi R \, dR,
    \label{eq:fdisk}
\end{equation}
where $F_{\lambda}$ is the observed spectral flux density at wavelength  $\lambda$, $d$ is the distance to the target, and $B_{\lambda}(T)$ is the Planck function in wavelength space. Using $E(B-V)$ and the extinction law of \cite{1989ApJ...345..245C} with $\rm R_{V}=3.1$, we calculated flux corrections from the UV to IR wavelengths.

We used the Least Squares Minimization method implemented in the {\tt scipy.optimize.least\_squares} module to estimate the best-fit parameters. To determine the uncertainties of the best-fit parameters, we performed a Monte Carlo simulation with 1000 iterations. In each iteration, we randomly varied the input data, such as distances, observed fluxes, and extinction $E(B-V)$, within their respective uncertainties and re-estimated the model parameters.

To reduce the number of free parameters in the SED modeling, we fixed some of them based on physically motivated assumptions. We fixed the WD mass at a conservative value\footnote{This is similar to the average mass of WDs ($M_\textrm{WD} = 0.81 \pm 0.05 M_\odot$) in AM CVns determined by \cite{2022MNRAS.512.5440V}. \label{note:wdmass}} of $M_{WD}=0.8 M_\odot$, and assumed that the inner part of the accretion disk is equal to the WD radius, $R_{in}\approx R_{WD}$. To estimate the WD radius, we used the WD mass-radius relation from \cite{1988ApJ...332..193V}. Additionally, we fixed the inclination angle\footnote{ The inclination angles are unknown for our targets, except for ZTF J2007$-$0527 ($i =69^{\circ}\pm2$; see Appendix~\ref{app:LC}). For the SED modeling, we adopted $i =60^{\circ}$. Our results can be compared to \cite{2018A&A...620A.141R}, where a similar value was used.\label{note:angle}} at $i = 60^{\circ}$. The final fitted parameters of the accretion disk were $\dot{M}$ and $R_{out}$. We constrained the maximum value of $R_{out}$ to be equal to the Roche-lobe radius of the WD based on \citet{1983ApJ...268..368E} approximation. Assuming a maximum WD mass of $\sim 1.4\, M_\odot$ and minimum allowable donor mass of $\sim 0.01\, M_\odot$ for an orbital period of  $\sim 17$ minutes, we computed the maximum Roche-lobe radius of the WD to be $\sim 0.17\, R_\odot$. 

Fig.~\ref{fig:sed} shows the observed SEDs of ZTF J1840$-$1742, ZTF J2007$-$0527, and ZTF J2111+3158, along with the corresponding  best-fit accretion disk models. For illustrative purposes, we show the possible contributions of the WD (green dashed lines; $T_{\rm eff} = 16000$~K, $R_{\rm WD} = 0.01~R_{\odot}$) and a cold donor (gray dotted lines; $T_{\rm eff} = 3000$~K, $R_{\rm donor} = 0.07~R_{\odot}$) using simple blackbody models. Evolutionary models for ultracompact binaries \citep[e.g.,][]{2006ApJ...640..466B, 2021ApJ...923..125W} suggest that a WD can be even hotter, with a temperature closer to $T_{\rm eff} \sim 30000$~K at an orbital period of $\sim$17 minutes. In either scenario, the combined contribution of the WD and the donor to the observed SED is negligible compared to the accretion disk component alone; therefore, we did not include them in our analysis.

\begin{figure}[h]
    \centering
    \includegraphics[width=0.5\textwidth]{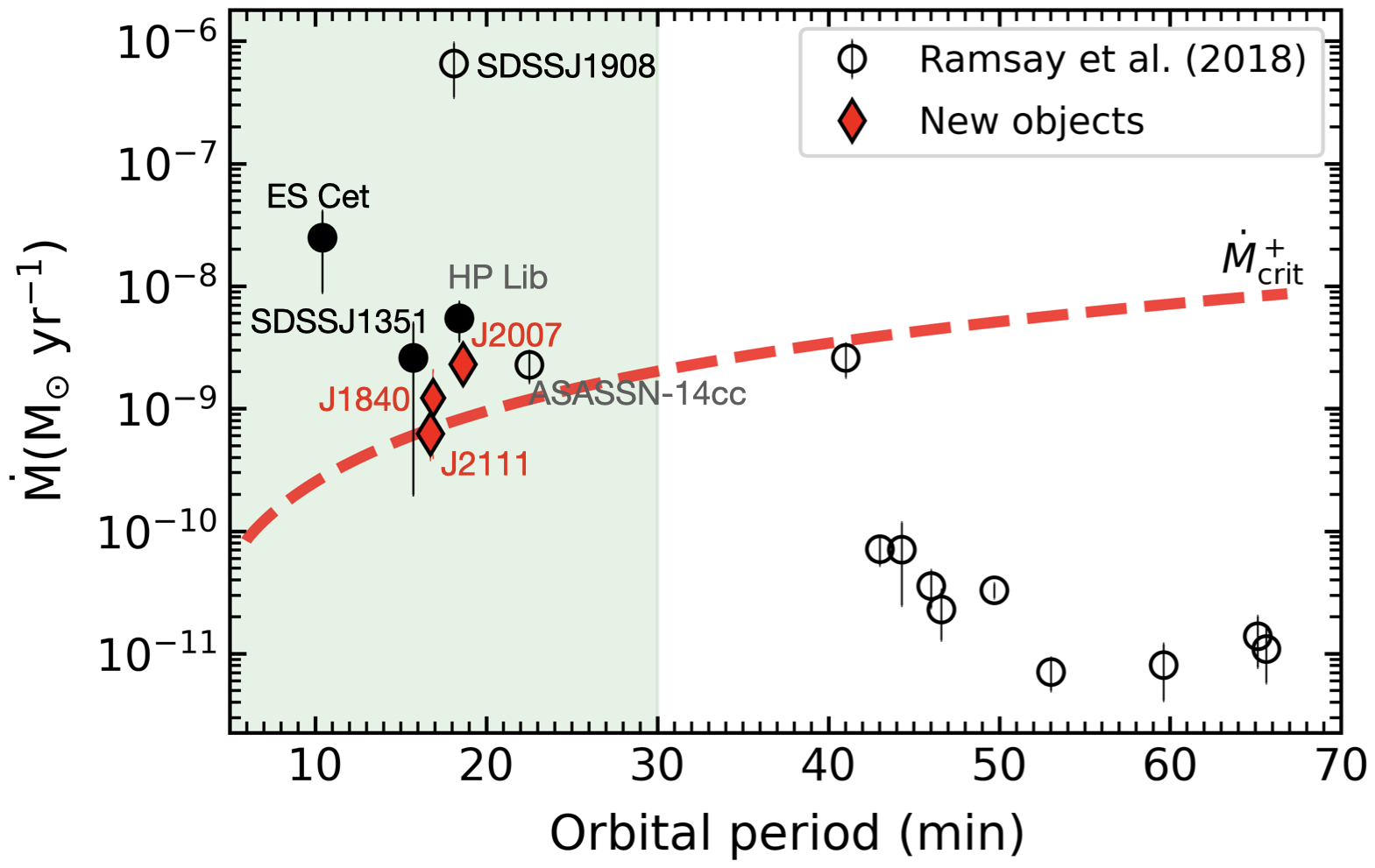}
    \caption{Orbital period vs. the mass accretion rate for AM~CVns. Open circles correspond to a sample of AM~CVns with measured mass accretion rates from \citet{2018A&A...620A.141R}. Our targets are shown as red diamonds. Black circles show known high-state AM~CVns recovered with our ZTF period search analysis. The green shaded region shows the orbital period range used to search for high-state AM~CVns. The red curve shows the critical mass-transfer rate ($\dot{M}_\mathrm{crit}^+$) above which a helium accretion disk is expected to remain stably hot (see Section~\ref{sec:population} and Eq.~(\ref{eq:Mdot_crit})). Our targets show mass accretion rates of the order of $ 10^{-9}\,M_{\odot}\,\mathrm{yr}^{-1}$, consistent with high-state AM~CVn systems.}
    \label{fig:mdot}
\end{figure}

Table \ref{tab:params} shows the  best-fit parameters of the SED modeling for our objects. The estimated $R_{out}$ values of our targets are consistent with the physically motivated boundaries of the WD's Roche lobe. We experimented with SED modeling and explored how varying the inclination angle\myfootref{note:angle} $i = (50-70)^{\circ}$ and the WD mass $M_{\rm WD}(M_{\odot})=(0.6-1.0)$ affects the  estimated accretion rate. This modeling shows that final mass accretion rates are stable across these ranges and agree with the values and uncertainties  in Table \ref{tab:params}. We did not include donor or WD  components in our SED modeling, as their combined contribution is only a few percent compared to the accretion disk model (see Fig.~\ref{fig:sed}). Fig.~\ref{fig:mdot} shows the mass accretion rate as a function of orbital period for a sample of AM CVn systems from \cite{2018A&A...620A.141R}. ZTF J1840$-$1742, ZTF J2007$-$0527, and ZTF J2111+3158 demonstrate mass accretion rates of the order of $10^{-9}\,M_{\odot}\,\mathrm{yr}^{-1}$ (see Table \ref{tab:params}), consistent with these systems being in a high state.

\subsection{X-ray luminosity}

We determined the $3\sigma$ upper limits on the X-ray flux for our objects using the XMM-Newton Upper Limit Server \citep{2020A&A...641A.136W}. We retrieved the observations containing the sky positions of ZTF~J1840$-$1742, ZTF~J2007$-$0527, and ZTF~J2111+3158, and then converted the count rates to X-ray fluxes in the 0.2--12 keV energy band by assuming a power-law spectral model with a photon index $\Gamma=2$. To account for interstellar absorption, we assumed hydrogen column densities of  $10^{20}\ {\rm cm}^{-2}$ for ZTF~J2111+3158, and  $10^{21}\ {\rm cm}^{-2}$ for ZTF~J1840$-$1742 and ZTF~J2007$-$0527,  consistent with their extinction values $E(B-V)$. We averaged the fluxes between all XMM-Newton upper flux measurements, and derived the X-ray luminosities based on Gaia distances. 
We estimated the $3\sigma$ upper limits on the X-ray luminosities in the 0.2--12 keV energy band for ZTF~J1840$-$1742, ZTF J2007$-$0527, and ZTF~J2111+3158 of $L_X\lesssim 2.1 \times 10^{33}$,  $\lesssim 5.2 \times 10^{33}$ and $\lesssim 2.0 \times 10^{33}$ $\rm erg\ s^{-1}$, respectively (see Table~\ref{tab:params}).

\begin{figure}[h]
    \centering
    \includegraphics[width=0.5\textwidth]{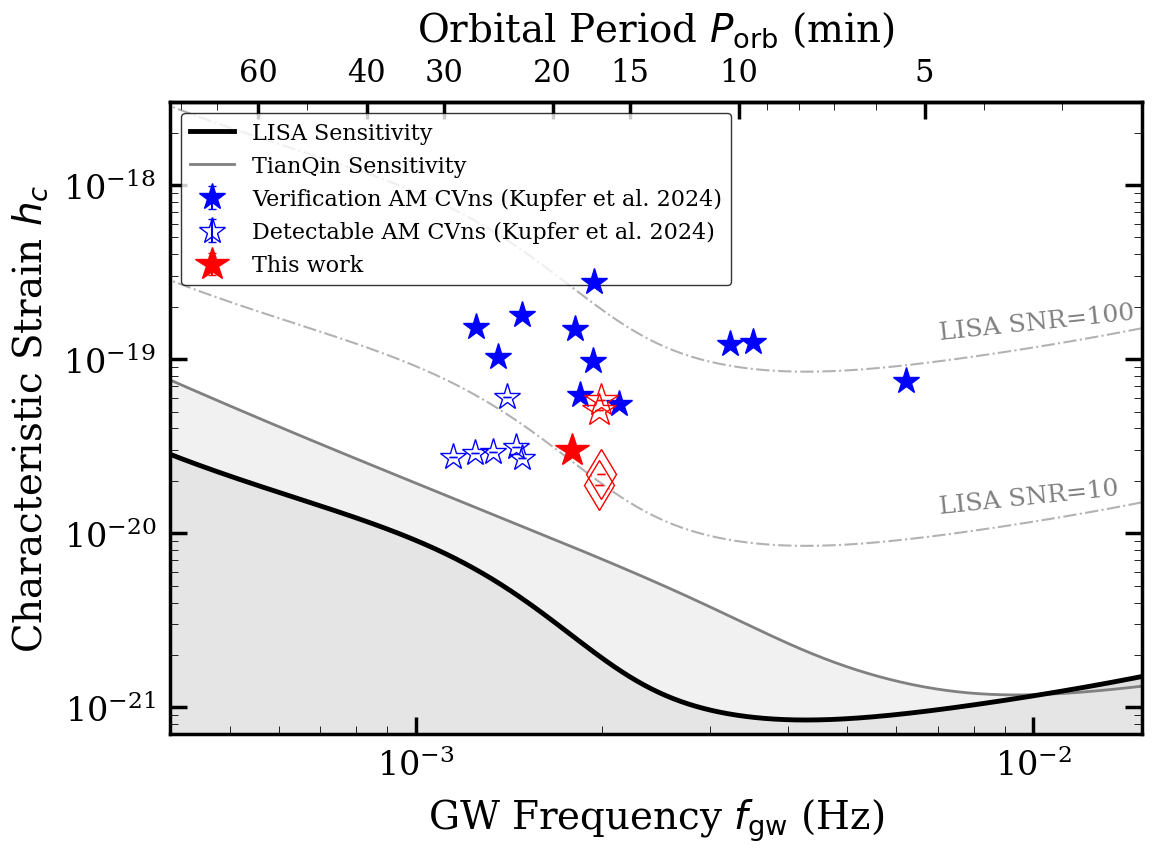}
    \caption{Characteristic GW strain for AM CVn binaries over a 4-year mission duration. Known LISA verification  (blue stars) and other detectable AM~CVn systems (open stars) are adopted from \citet[][see their Table~1]{2024ApJ...963..100K}. The sensitivity curves for LISA and TianQin are shown in black and gray, respectively. Red (filled) star shows  ZTF~J2007$-$0527 with a donor mass of $ M_{\rm donor}\approx0.07\,M_\odot$ and WD mass of $ M_{\rm WD}\approx0.6\,M_\odot$. The open symbols correspond to ZTF~J1840$-$1742 and ZTF~J2111+3158, showing lower and upper donor masses of $ M_{\rm donor}\approx 0.03\,M_\odot$ (red diamond) and $\approx0.1\,M_\odot$ (red stars), respectively, both assuming a WD mass of $M_{\rm WD}\approx0.8\,M_\odot$. ZTF~J1840$-$1742,  ZTF~J2007$-$0527 and ZTF~J2111+3158 are expected to be detectable sources for both LISA and TianQin over 4-year missions. }
    \label{fig:gw}
\end{figure}

\section{Discussion}
\label{sec:discussion}

\subsection{Properties of the new objects}

\subsubsection{Expected donor masses}
\label{sec:donor}

We theoretically constrained the donor masses of our targets using (i) the period-density relation and (ii)  evolutionary models for the three AM CVn formation channels. We also modeled the light-curve of the ZTF~J2007$-$0527 and compared it with theoretical values (see Appendix~\ref{app:LC}).

(i) The period-density relation is given by: 
\begin{gather}
   \overline{\rho}_{\rm donor}= \frac{3M_{\rm donor}}{4\pi R_{\rm donor}^3} \approx 107\,{\rm g\,cm^{-3}}\left(P_{\rm orb}/{\rm hr}\right)^{-2},
    \label{eq:pho_donor}
\end{gather}
where $\overline{\rho}_{\rm donor}$ is the donor's mean density, and $M_{\rm donor}$ and $R_{\rm donor}$ are the mass and radius of the donor. We assumed the zero temperature WD mass-radius relation from \cite{1988ApJ...332..193V} to estimate the lower limit of the donor masses from Eq.~(\ref{eq:pho_donor}). For the orbital periods of 16.71, 16.86 and 18.63 minutes, these resulted in donor masses of 0.037, 0.036 and 0.032 $M_\odot$, respectively.

(ii) We considered the following evolutionary channels for AM CVns. For the evolved-CV channel, we adopted the model of \citealt{2023A&A...678A..34B} (see their Fig.~1 and model~3) with initial parameters $M_1=0.8\,M_\odot,\ M_2=1.1\,M_\odot,\ P_\mathrm{orb}=2.60\,$d. For the WD channel, we adopted the model of \citealt{2021ApJ...923..125W} with initial masses of $M_1=0.75\,M_\odot,\ M_2=0.18\,M_\odot$ and initial central donor specific entropy of $\rm S_{c, He}/N_Ak_B=3.07$, where $N_A$ is Avogadro's number. For the helium star donor channel, we used the model of \citealt{2015ApJ...807...74B} (see also \citealt{2021ApJ...923..125W}), with initial parameters $M_1=0.5\,M_\odot,\ M_2=0.35\,M_\odot,\ P_{\rm orb}=20$ minutes. For the given orbital periods of our targets, the WD and evolved CV channels provide the donor masses of $ M_{\rm donor}\approx0.04\ M_\odot$. The helium star channel gives the donor mass around $ M_{\rm donor}\approx0.1\ M_\odot$. 

These two methods suggest that the donor masses of our targets are within the $M_{\rm donor}\approx 0.03-0.1\ M_\odot$ range. From the light curve modeling we independently estimated the donor mass of ZTF~J2007$-$0527 to be $M_{\rm donor}\approx0.07\ M_\odot$ (see Appendix~\ref{app:LC} and Fig.~\ref{fig:J2007_fit}). This value is consistent with the theoretical  predictions. The donor mass of the ZTF~J2007$-$0527  suggests that the system formed via the helium star channel, rather than WD or evolved CV channels.

\subsubsection{Gravitational wave emission}
\label{sec:gw}

To compute the characteristic strain ($h_c$) for our targets,  we first calculated the dimensionless GW amplitude ($\mathcal{A}$), which is given by \cite{2012A&A...544A.153S} as:
\begin{gather}
   \mathcal{A} = \frac{4(G\mathcal{M})^{5/3}}{c^4 d }(\pi f_{gw})^{2/3}
    \label{eq:A}
\end{gather}
where $\mathcal{M} = (M_1 M_2)^{3/5}/ (M_1+M_2)^{1/5}$ is the chirp mass, and  $M_1$ and $M_2$ are the component masses,  $f_{gw} = 2/P_{\rm orb}$ is the GW frequency. The characteristic strain for our targets is computed as $h_c = \sqrt{N_{\rm cycle}} \mathcal{A}$, where $N_{\rm cycle} = f_{\rm gw} T_{\rm obs}$ and $T_{\rm obs}$ is the observation time.

Fig.~\ref{fig:gw} shows the characteristic strain of our targets along with LISA verification and detectable AM CVns \citep[][see their Table 1]{2024ApJ...963..100K}.  The sensitivity curves for LISA and TianQin were computed using the {\tt legwork} \citep{2022ApJS..260...52W} for $T_{\rm obs} = 4$ years.  For ZTF~J2007$-$0527, we used the donor mass of  $M_{\rm donor}\approx0.07\,M_\odot$ and WD mass of $M_{\rm WD}\approx0.6\,M_\odot$ estimated from the light-curve modeling (see Appendix~\ref{app:LC}). For ZTF~J1840$-$1742 and ZTF~J2111+3158, we assumed the lower and upper limits of the donor masses to be $ M_{\rm donor}\approx 0.03\,M_{\odot}$ and  $0.1\,M_{\odot}$, respectively, both assuming a WD mass of $M_{\rm WD}\approx0.8\,M_\odot$. LISA and TianQin are expected to detect ZTF~J1840$-$1742, ZTF~J2007$-$0527, and ZTF~J2111+3158 as GW sources over their 4-year missions.

\subsection{High-state AM CVn population in the Milky Way}
\label{sec:population}

\subsubsection{Local space density}

Using our target selection criteria, we identified three new high-state AM~CVn binaries and recovered three previously known systems (see Appendix~\ref{app:list}). Due to the small sample size,  we estimated the order of magnitude of the local space density for high-state AM~CVns using the $1/V_{\rm max}$ method \citep{1968ApJ...151..393S}. As the distances of the six objects vary from approximately 280~pc up to 1800~pc, we assumed a local volume where the density follows only a vertical distribution \citep{1993ApJ...414..254T}:
\begin{gather}
V_{max,i} = \Omega \int_{0}^{d_{max,i}}  e^{-(|r\ {\rm sin}b_i|)/h_z} r^2 dr\ ,
\label{eq:vmax}
\end{gather}
where $\Omega$ is the solid angle covered by the survey ($4\pi$ for an all-sky survey), $h_z$ is the Galactic disk scale height, $b_i$ is the Galactic latitude of the $i$-th source, and $d_{max, i}$ is the maximum distance at which the source would remain detectable by the survey sensitivity and selection criteria.

In target selection, we defined the Gaia sample using multiple criteria, based on quality cuts on magnitude, parallax  and proper motions: $G<21$ mag; $\varpi > 0.1$, $\varpi/\sigma_{\varpi} \ge 2$; $\mu_{\alpha}/\sigma_{\mu\alpha*} \ge 5$ and $\mu_{\delta}/\sigma_{\mu\delta} \ge 5$ (see Section~\ref{sec:survey}). These quality cuts remove the distant sources from the sample, even if they satisfy the magnitude cut of $G<21$ mag. To account for the volume change due to these quality cuts,  we computed for each of our targets the magnitude-limited distance ($d_{\rm mag}$), the parallax-limited distance ($d_{\rm \varpi}$), and the proper motion-limited distance ($d_{\rm \mu}$). We defined the final maximum distance ($d_{max}$) as the minimum of these distances and the survey distance limit, $d_{max} = min(d_{\rm mag};d_{\rm \varpi}; d_{\rm \mu}; 10\ {\rm kpc})$.

For any distance $d$, the apparent magnitude $G$ is given by:
\begin{gather}
G = M_{G,0} + 5\log_{10}(d) - 5 + A_G(d, l, b),
\label{eq:gmag}
\end{gather}
where the extinction along the line of sight, $A_G(d, l, b)$, depends on distance $d$ and Galactic latitude ($b$) and longitude ($l$). We computed the $d_{\rm mag}$ using $G=21$ mag limit, the extinction-corrected absolute magnitude $M_{G,0}$ of the targets and the $A_G(d, l, b)$ from the {\tt mwdust} package ({\tt Combined19map}). To estimate  $d_{\rm \varpi}$ and $d_{\rm \mu}$, we used the median standard uncertainties $\sigma_{\varpi}(G)$, $\sigma_{\mu\delta}(G)$ and $\sigma_{\mu\alpha*}(G)$ as a function of magnitude for stars from Gaia EDR3 \citep[see Table 4 of][]{2021A&A...649A...2L}. We calculated the $d_{\rm \varpi}$ by solving the equation $\varpi(d)/\sigma_{\varpi} (G)=2$, where $G$ is given by Eq.~(\ref{eq:gmag}) and $\varpi(d)=1000/d$~mas. To estimate $d_{\rm \mu}$, we first calculated the transverse velocity ($v_{trans}$) for each of our targets at their distances, and then derived the proper motion as $\mu(d) = v_{trans}/ (4.74\times d)$.  We solved the equation $\mu(d)/\sigma_{\mu} (G)=5$ to estimate the $d_{\rm \mu}$, where the total proper-motion standard uncertainty is defined as $\sigma_{\mu}(G)^2$$=\sigma_{\mu\alpha*}(G)^2 + \sigma_{\mu\delta}(G)^2$.

Using Eq.~(\ref{eq:vmax}), we computed the maximum volume for each of the six targets. We corrected the maximum volume for the following factors:

{\it (a) Gaia completeness.} For each target, we computed the completeness ($f_{Gaia, i}$) at $G=21$~mag using the empirical \textit{Gaia}~DR3 selection function based on the \texttt{gaiaunlimited} module \citep{2023A&A...669A..55C}.

{\it (b) ZTF footprint and geometric correction.}  We accounted for the fraction of the sky covered by ZTF ($\delta \geq -28^{\circ}$) with number of epochs $N > 50$. This resulted in a correction factor of $f_{ZTF}\approx0.73$ (or $\approx73\%$), corresponding to the effective area of the ZTF survey footprint. We also accounted for both the recovery fraction of periodic signals in our ZTF period search analysis and the geometric selection effects resulting from the random orientation of AM~CVn orbits (see Appendix~\ref{app:ztfrecovey} for details). We determined the total detection probability to be $ \approx96.5-99.8\%$ for our targets, corresponding to  a correction factor of $f_{det,i}\approx0.965$ for the volume.

{\it (c) Correction for low-amplitude variable systems.} Our simulations in Appendix~\ref{app:ztfrecovey} show that our ZTF period search analysis is highly complete for systems with intrinsic amplitude $\ga 0.05$ mag, which resulted in the identification of six targets (three known and three new systems). However, two known objects, AM CVn and SDSS J1908+3940, passed our Gaia selection criteria and lie within the ZTF footprint (see Appendix~\ref{app:list}), but showed no detectable variability in their ZTF light curves, implying that their intrinsic amplitude is $\la 0.05$ mag. This shows that our survey is incomplete for systems with intrinsically low photometric variability amplitudes.  Among five known systems passing the Gaia cuts and lying within the ZTF footprint, we detected variability for three systems (ES Cet, SDSS J1351$-$0643, HP Lib) and missed two objects (AM CVn and SDSS J1908+3940). The fraction $f_{A}\sim 3/5$ represents the empirical completeness of our survey of the full high-state AM CVn population, accounting for the presence of low-amplitude variable systems that are "invisible" to our ZTF period search analysis.  

We accounted for all the correction factors above by defining a corrected maximum volume for each system as:
\begin{gather}
V_{max,i}^{\rm corr} = (V_{max,i} \times f_{Gaia, i} \times f_{det,i}) \times f_{ZTF} \times f_{A}.
\label{eq:Vcorr}
\end{gather}

The local space density ($\rho_0$) and its uncertainty ($\sigma_{\rho_0}$) are computed as follows:

\begin{gather}
\rho_0 = \sum_{i=1}^{6} \frac{1}{V_{max,i}^{\rm corr}}\quad \text{and} \quad {\rm \sigma_{\rho_0}^2} =  \sum_{i=1}^{6} \frac{1}{(V_{max,i}^{\rm corr})^2}.
\label{eq:rho}
\end{gather}

Using Eqs.(\ref{eq:vmax})--(\ref{eq:rho}), we estimated the local space density of the high-state AM~CVn population and its uncertainty to be $\rho_0 = (1.0\pm 0.5) \times 10^{-8}~\text{pc}^{-3}$, assuming a disk scale height of $h_z = 300$~pc. Assuming a smaller disk scale height of $h_z = 200$~pc, we obtained a space density of  $\rho_0 = (2.7 \pm 1.5) \times 10^{-8}~\text{pc}^{-3}$, which is higher by a factor of $\approx 3$ than the value for $h_z = 300$~pc. Our estimates of the local space density suggest that high-state systems account for $\approx 2-5\%$ of the total AM~CVn population density \citep[e.g., $\rho_0=(5.5 \pm 3.7) \times 10^{-7}~\text{pc}^{-3}$;][]{2025PASP..137a4201R}.

We note that our high-state AM CVn selection box on the Gaia color-color diagram is defined empirically based on the small number of previously known systems (see Section~\ref{sec:survey}). Such target selection could bias our search towards systems similar to already identified high-state AM CVns. We doubled the size of our search box on the Gaia color-color diagram and still found no additional systems. This suggests that our initial selection criteria are highly complete for systems with SEDs similar to known high-state AM CVns.

\subsubsection{Total number of objects}

We computed the total number of high-state AM~CVn systems in the Milky Way based on the local space density, $\rho_0$. We modeled the density distribution as a double exponential disk:
\begin{gather}
\rho(R,z) =  \rho_0 \exp\left( -\frac{R - R_{\rm Sun}}{h_R} \right) \exp\left( -\frac{|z|}{h_z} \right) ,
\label{eq:rho_model}
\end{gather}
where $R_{\rm Sun}=8$~kpc is the distance from the Galactic Center to the Solar neighborhood, $h_R=3$~kpc is the radial scale length. The total number of objects is given by:
\begin{gather}
N_{\rm total} = \int_{0}^{2\pi} \int_{0}^{R_{\rm max}} \int_{-z_{\rm max}}^{z_{\rm max}} \rho(R, z) R \, dR \, dz \, d\phi,
\label{eq:ntotal}
\end{gather}
where a maximum Galactic disk radius of $R_{\rm max}=16$~kpc and a vertical height of $z_{\rm max}=5 \times h_z$ were adopted. 

Table~\ref{tab:amcvn_detections} presents the total number of high-state AM CVn binaries in the Milky Way, computed for a Galactic disk model with $h_z = 200$ and $300$~pc, alongside the lower, median, and upper limits of the local space density. For a space density of  $\rho_0 = (2.7 \pm 1.5) \times 10^{-8}~\text{pc}^{-3}$ ($h_z = 200$~pc), the calculation yields a total number of $N_{\rm total}\approx 3910-13\ 330$ high-state AM CVns in the Milky Way, while for $\rho_0 = (1.0\pm 0.5) \times 10^{-8}~\text{pc}^{-3}$ ($h_z = 300$~pc) we estimated $N_{\rm total}\approx 2350-7150$ systems. For a space density of all AM CVns of $\rho_0 = (5.5 \pm 3.7) \times 10^{-7}~\text{pc}^{-3}$ \citep[$h_z \approx 200$~pc;][]{2025PASP..137a4201R}, the calculation yields a total of $N_{\rm total}\approx 56\ 400-288\ 400$  objects in the Milky Way.

\begin{figure}[h!]
    \centering
    \includegraphics[width = 1.0\linewidth]{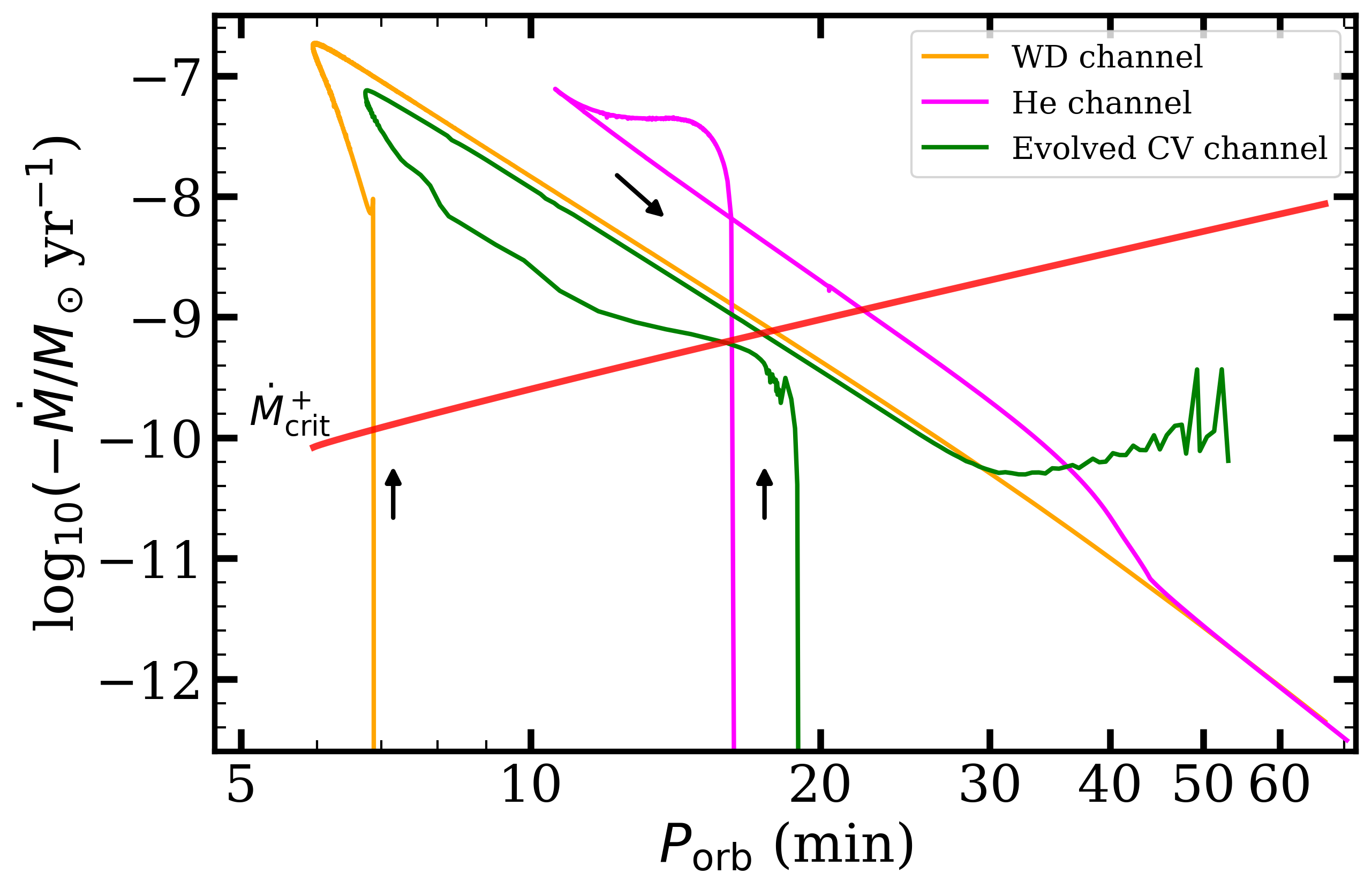}
    \includegraphics[width = 1.0\linewidth]{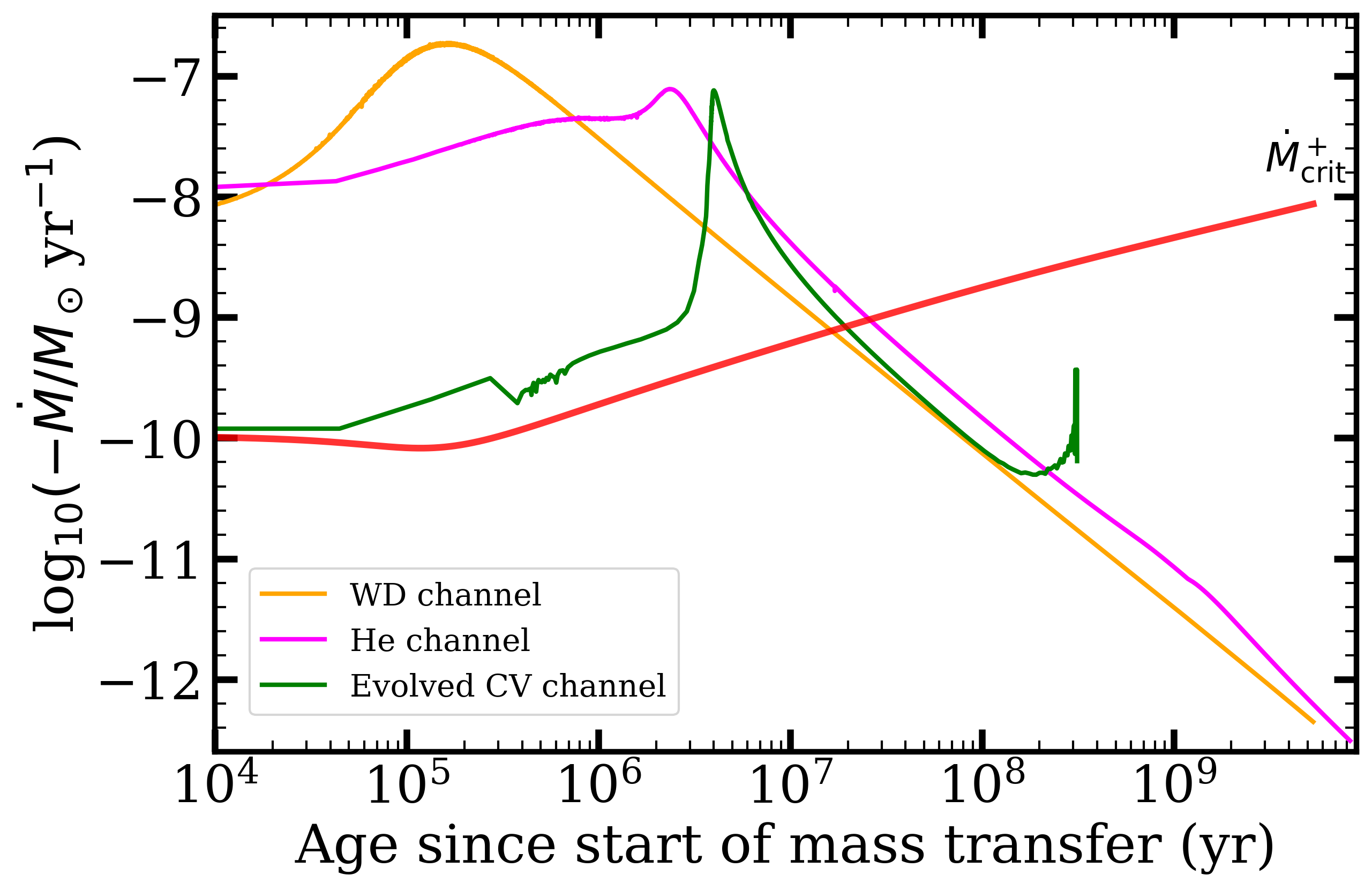}
    \caption{Mass-transfer rate as a function of orbital period (top) and time since the onset of mass transfer (bottom). Three formation channels are shown: the WD channel (orange; \citealt{2021ApJ...923..125W}), the helium star channel (magenta; \citealt{2015ApJ...807...74B, 2021ApJ...923..125W}), and the evolved CV channel (green; \citealt{2023A&A...678A..34B}). The red curve in both panels marks the critical mass-transfer rate ($\dot{M}_\mathrm{crit}^+$) above which a helium accretion disk is expected to remain stably hot. Systems with a mass transfer rate exceeding the critical rate are classified as high-state AM~CVns.}
    \label{fig:models}
\end{figure}

\subsubsection{Characteristic lifetime and birth rate}

To determine the characteristic lifetime ($\tau_{\rm life}$) of AM CVns, we used representative evolutionary tracks for the three proposed formation channels (see Section~\ref{sec:donor} for details). Fig.~\ref{fig:models} (top panel) shows  evolutionary tracks for the three evolutionary channels of AM CVns in the $\dot{M}$--$P_{orb}$ plane. We define a system as being in a high-state when its mass accretion rate exceeds the critical threshold required to maintain a hot and stable helium accretion disc\footnote{A subset of ultra-compact binaries with $P_{\rm orb} \lesssim 10$ minutes are known to undergo direct-impact accretion \citep[e.g.,][]{2004MNRAS.350..113M}. 
Due to their fast evolution, we computed characteristic lifetimes only for high-state AM CVn systems having a stable helium accretion disk.}. 
The critical mass-transfer rate ($\dot{M}_{\rm crit}^{+}$) is calculated following the models of \citealt{2012A&A...544A..13K} for a chemical composition of $Y=0.98, Z=0.02$ as follows:
\begin{gather}
    \dot{M}_{\mathrm{crit}}^{+} = \dot{M}_0\,
    \left(\frac{\alpha_\mathrm{visc}}{0.1}\right)^{-0.05}
    \left(\frac{R_\mathrm{out}}{R_\odot}\right)^{2.67}
    \left(\frac{M_\mathrm{WD}}{M_\odot}\right)^{-0.89},
    \label{eq:Mdot_crit}
\end{gather}
where $\dot{M}_0=1.29\times10^{-7}\,M_\odot\,\mathrm{yr}^{-1}$, $\alpha_\mathrm{visc}$ is the viscosity parameter of the accretion disk, $R_\mathrm{out}$ is the outer radius of the disk, and $M_\mathrm{WD}$ is the WD mass. The dependence on $\alpha_\mathrm{visc}$ is negligible and we used a typical value of $\alpha_\mathrm{visc}=0.1$. The outer disk radius can be approximated by the tidal truncation radius, $R_\mathrm{T}$, beyond which tidal perturbations prevent a steady disk from extending further \citep{2003cvs..book.....W}:
\begin{gather}
    R_\mathrm{out}\approx R_\mathrm{T}=a\frac{0.6}{1+q},
    \label{eq:R_out}
\end{gather}
where $a$ is the separation between the stars and $q=M_2/M_1$ is the mass ratio. We computed $\dot{M}_{\rm crit}^{+}$using Eq.~(\ref{eq:Mdot_crit}) and (\ref{eq:R_out}) and the binary parameters from each evolutionary model. Since the differences between the various models are negligible, we used $\dot{M}_{\rm crit}^{+}$ only for the WD channel as a representative threshold (see red lines in Figs.~\ref{fig:models}). 

Fig.~\ref{fig:models} (bottom panel) shows the mass accretion rate as a function of time since the onset of mass transfer for three evolutionary channels. In all three formation channels, the systems enter the high-state ($\dot{M}\gtrsim\dot{M}_\mathrm{crit}^{+}$) at an early phase, soon after accretion has started. The mass transfer rate rises to a maximum of $\dot{M}\approx10^{-7}\,M_\odot\,\mathrm{yr}^{-1}$, after which the binaries reach the period minimum, and reverse their evolution toward longer periods with declining $\dot{M}$. Systems evolving via the WD and evolved CV channels exit the high-state at very similar orbital periods of approximately $P_{orb}\approx18\,$ minutes, whereas those following the helium star channel exit at the slightly later period of $P_{orb}\approx 22$ minutes. The characteristic lifetime of these systems being in high-state is about $\tau_{\rm life}\approx20\,$Myr, with a maximum variation factor of $\sim$1.3 between different evolutionary channels. To define the total lifetime of AM CVns, we consider the evolution up to orbital periods of $P_{orb}\approx60\,$ minutes at which point the mass accretion rate drops to $\dot{M}\lesssim10^{-12}\,M_\odot\,\mathrm{yr}^{-1}$. This corresponds to characteristic total lifetimes for AM~CVn systems of $\tau_{\rm life}\approx4\,$Gyr, with a variation of a factor of $\sim$1.3 between the helium star and WD channels\footnote{For the evolved CV channel, numerical simulations stopped when either the donor mass dropped to $M_{2}\approx 0.01\,M_{\odot }$, or the donor age exceeded the Hubble time \citep{2023A&A...678A..34B}. CVs evolving via this channel first undergo a phase of mass transfer at longer periods typical of hydrogen-rich CVs, so their total evolutionary age at this point is much longer compared to other AM~CVn channels ($\approx 11$~Gyr). Therefore, we limited our estimates of the total characteristic AM~CVn lifetimes to the WD and helium star channels, which could be tracked consistently to the $\approx 60\,$ minutes period threshold.}. We note that there is an uncertainty in the total lifetime due to the poorly constrained donor cooling at long orbital periods. Observed long-period AM~CVn systems appear difficult to explain via  cool accretor models \citep{2018A&A...620A.141R,2022MNRAS.512.5440V}, although this comparison may be affected by residual disk or boundary-layer emission (see \citealt{2021ApJ...923..125W} for a discussion). Therefore, the total lifetime may be uncertain toward shorter values by a factor of a few. Also, there may be additional AM CVn evolutionary models with different magnetic braking prescriptions, mass and period distributions, etc. but they are not considered in this work and should be explored further in the future.

We computed the birth rate of AM~CVns in the Milky Way by dividing the total number of systems by the characteristic lifetime:
\begin{gather}
R_{\text{birth}}= \frac{N_{\text{total}}}{\tau_{\text{life}}}.
    \label{eq:brate}
\end{gather}

For high-state AM~CVn binaries, this results in a birth rate of $R_{\text{birth}}\approx(4.3\pm2.4)\times 10^{-4}~\text{yr}^{-1}$ ($h_z = 200$~pc) and $R_{\text{birth}}\approx(2.4\pm1.2)\times 10^{-4}~\text{yr}^{-1}$ ($h_z = 300$~pc). For the total population of AM~CVns the birth rate is $R_{\text{birth}}\approx (4.3\pm2.9)\times 10^{-4}~\text{yr}^{-1}$ \citep[for $h_z \approx 200$~pc and $\rho_0 = (5.5 \pm 3.7) \times 10^{-7}~\text{pc}^{-3}$;][]{2025PASP..137a4201R}. These values are statistically consistent at the $1\sigma$ level, implying that most AM~CVn systems entering the high-state survive and reach the long-period low-state ($P_{orb}\gtrsim 30\,$ minutes). However, the large error bars prevent a definitive conclusion; therefore, we cannot rule out that some systems fail to survive the high-state phase, potentially leading to stellar mergers or thermonuclear detonations, such as the predicted Type .Ia supernovae \citep{2007ApJ...662L..95B}.

In comparison to binary population synthesis models, \citet{2001A&A...368..939N} predict that birth rates in the Milky Way are $(0.04-4.7) \times 10^{-3}\,\text{yr}^{-1}$ for the WD channel and $(0.9-1.6) \times 10^{-3}\,\text{yr}^{-1}$ for the helium star channel. Such birth rates yield a total number of objects of $(0.02-4.9) \times 10^{7}$ and $(1.8-3.1) \times 10^{7}$ for the WD and helium star channels, respectively, with a total space density of $(0.4-1.7) \times 10^{-4}\,\text{pc}^{-3}$. For the helium star channel, \citet{2022A&A...668A..80L} predict a birth rate of $4.6 \times 10^{-4}\,\text{yr}^{-1}$ with a total of $112\ 000$ objects with orbital periods $P_{orb} < 42-43$\,minutes, yielding a local space density of $3.1 \times 10^{-8}\,\text{pc}^{-3}$. For the evolved CV channel, \citet{2003MNRAS.340.1214P} predict a birth rate of $(0.5-1.3) \times 10^{-3}\,\text{yr}^{-1}$. In contrast to our study, these models simulate the entire AM~CVn population without focusing on high-state systems. However, even for the entire population, these models consistently overpredict the number of observed AM~CVn binaries, but provide the birth rate that is consistent in order of magnitude with the observed AM CVn population.

\begin{table}[ht]
\centering
\caption{Median number of high-state AM CVn systems expected to be detectable with LISA and TianQin during a 4-year missions (with a signal-to-noise ratio $(S/N) \ge 5$), and 10-year survey with LSST (at a limit of $m_{r}\lesssim26.9$~mag). The total and detectable numbers of sources are computed for lower, mean and upper values of the local space densities of $\rho_0 = (2.7 \pm 1.5) \times 10^{-8}~\text{pc}^{-3}$ ($h_z=200$~pc) and $\rho_0 = (1.0\pm 0.5) \times 10^{-8}~\text{pc}^{-3}$ ($h_z=300$~pc). } 

\label{tab:amcvn_detections}
\renewcommand{\arraystretch}{1.1} 
\setlength{\tabcolsep}{2pt}   
\begin{tabular*}{\columnwidth}{@{\extracolsep{\fill}} c|c|cccc} 
\hline
\hline
Disk scale  & Space density, $\rho_0$ & $N_{\rm total}$ & $N_{\rm LISA}$ & $N_{\rm TianQin}$ & $N_{\rm LSST}$  \\
height, $h_z$ & ($\times 10^{-8}\ \text{pc}^{-3}$) &  &  &  &  \\
\hline

\multirow{3}{*}{200 pc} & 1.2 & 3909 & 1411 & 758 & 1347 \\
                        & 2.7 & 8617 & 3108 & 1671 & 2970 \\
                        & 4.2 & 13326 & 4809 & 2585 & 4610 \\
\hline

\multirow{3}{*}{300 pc} & 0.5 & 2352 & 848 & 456 & 1023 \\
                        & 1.0 & 4749 & 1712 & 920 & 2071 \\
                        & 1.5 & 7146 & 2578 & 1384 & 3114 \\
\hline

\hline
\end{tabular*}
\end{table}

\begin{figure*}
    \centering
    \includegraphics[width=0.49\linewidth]{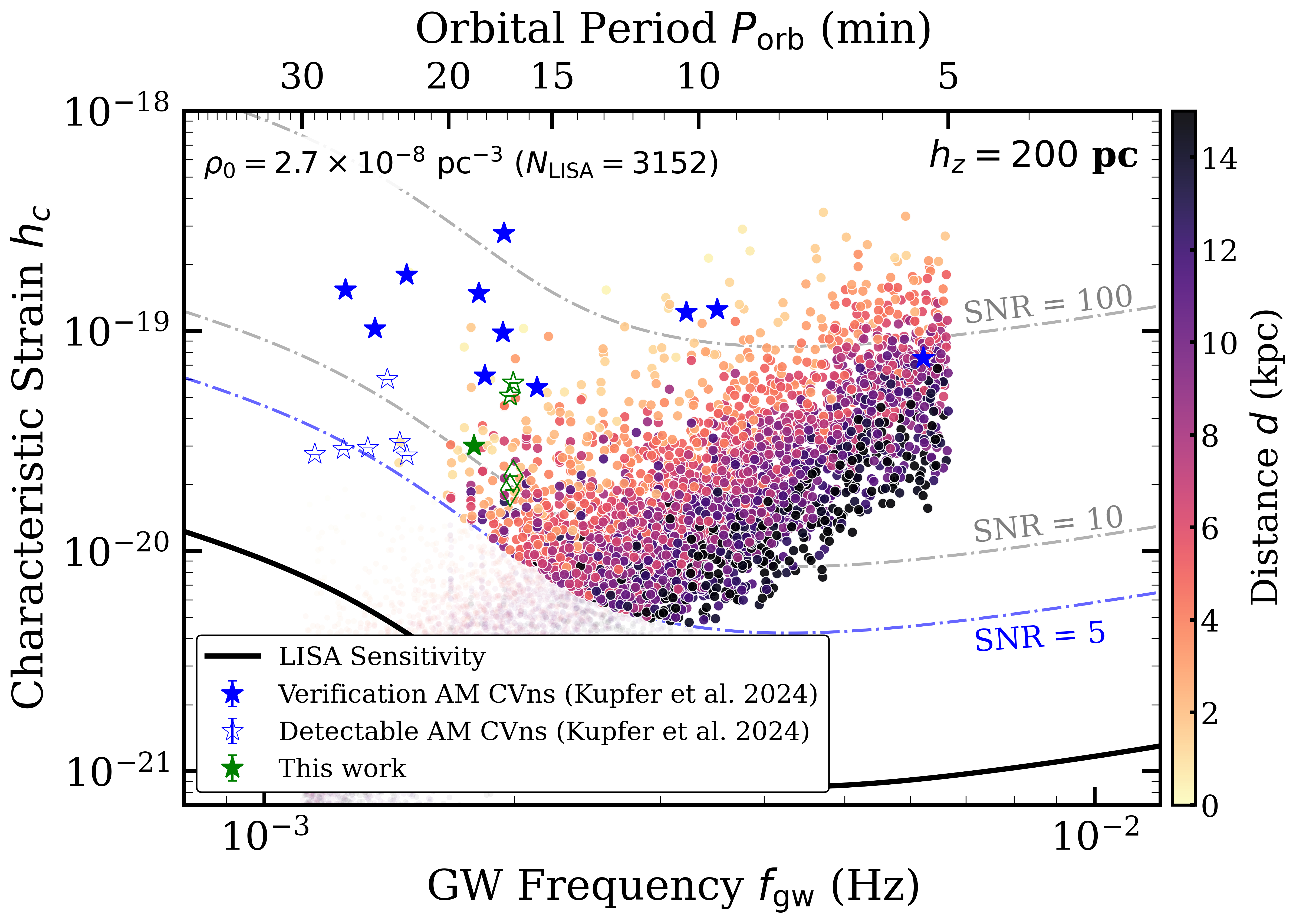} 
    \includegraphics[width=0.49\linewidth]{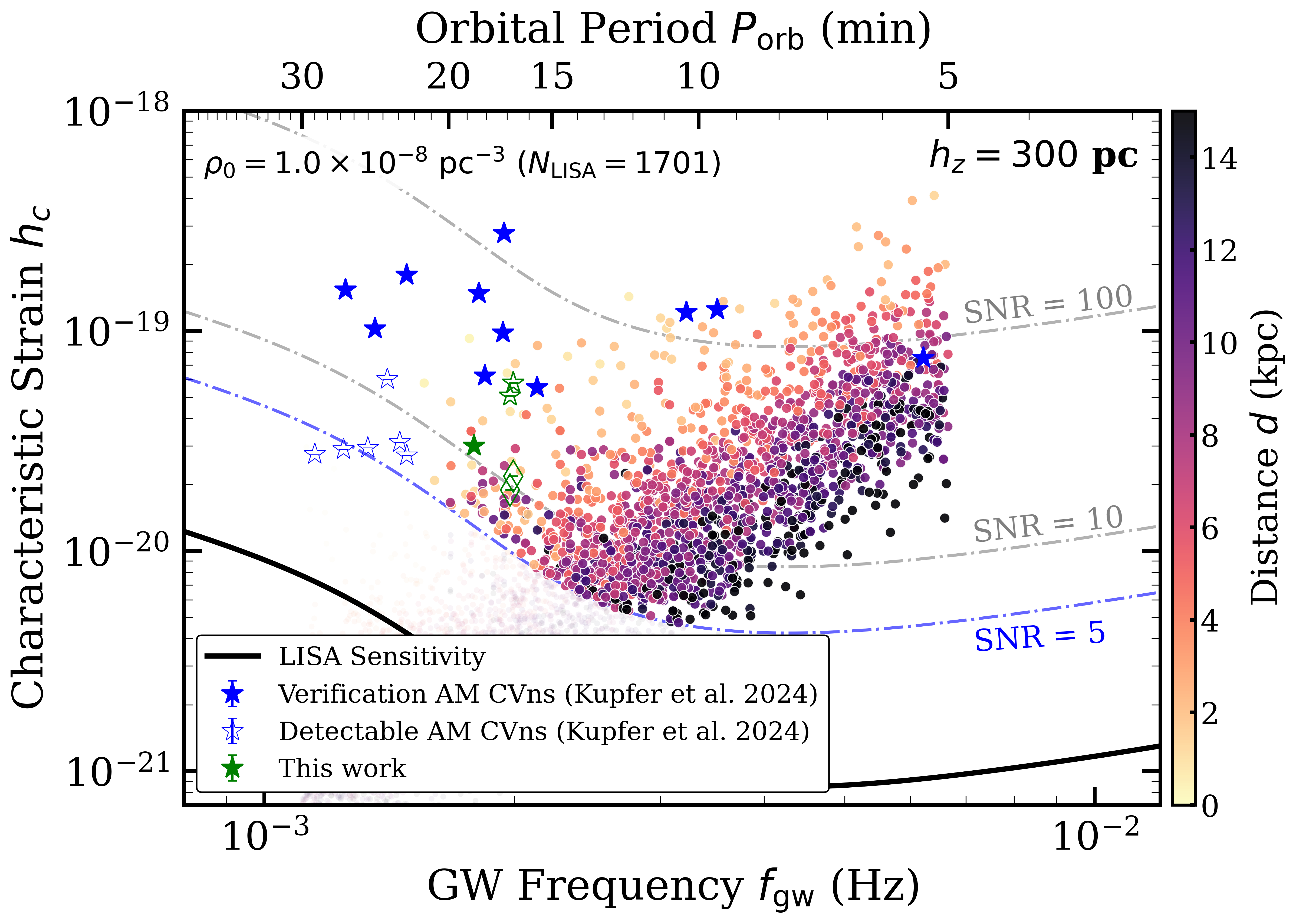}
    \includegraphics[width=0.49\linewidth]{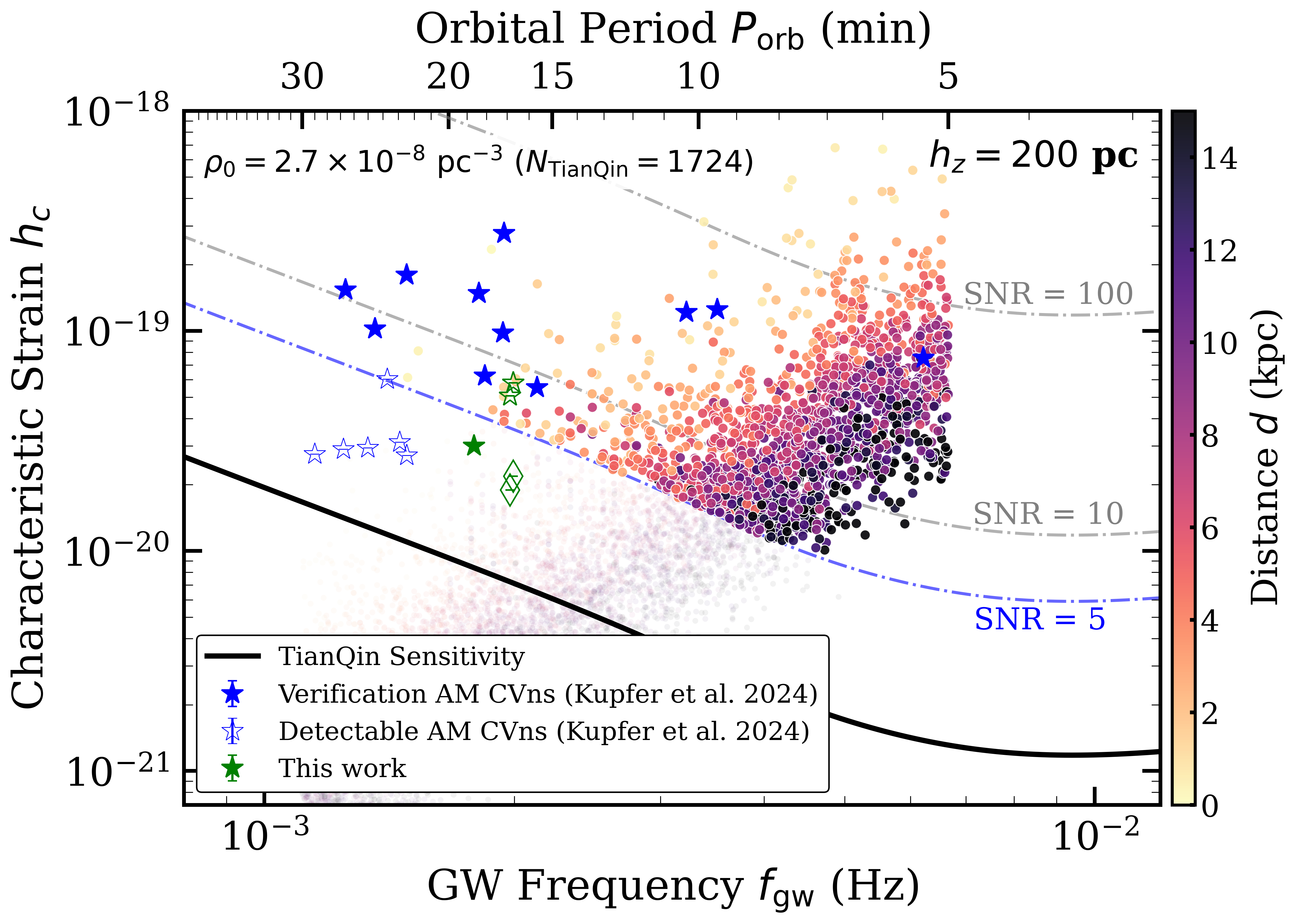} 
    \includegraphics[width=0.49\linewidth]{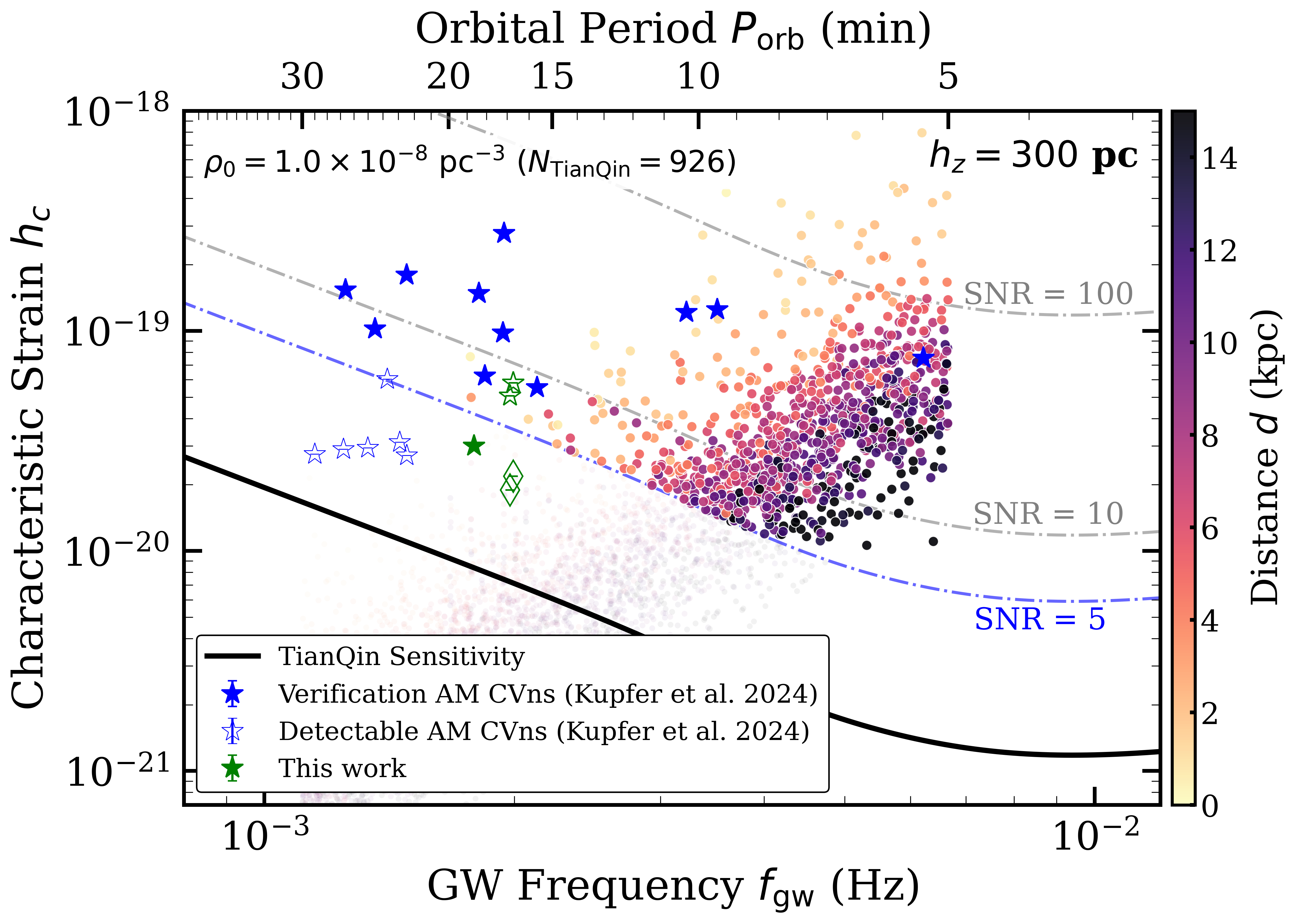}
    \caption{ Detectability of the high-state AM CVn population in the Milky Way with LISA (top) and TianQin (bottom). The  number of detectable systems is estimated based on a local space density of  $\rho_0=2.7 \times 10^{-8}\ \text{pc}^{-3}$ ($h_z=200$~pc, left panels) and $\rho_0=1.0 \times 10^{-8}\ \text{pc}^{-3}$ ($h_z=300$~pc, right panels) for a single realization.  The color bars show distance to the objects. Known LISA verification (blue stars) and other detectable AM~CVn systems (open stars) are adopted from \citet[][see their Table~1]{2024ApJ...963..100K}. New systems identified in this study are shown with green stars and diamonds. The solid lines show the sensitivity curves for LISA and TianQin, while the dashed lines correspond to signal-to-noise ratio of 5, 10 and 100. About 36\% of the total population is expected to be detectable by LISA and 19\% by TianQin during a 4-year mission with a signal-to-noise ratio $(S/N) \ge 5$.}
    \label{fig:GWpopulation}
\end{figure*}

\begin{figure*}
    \centering
    \includegraphics[width=0.49\linewidth]{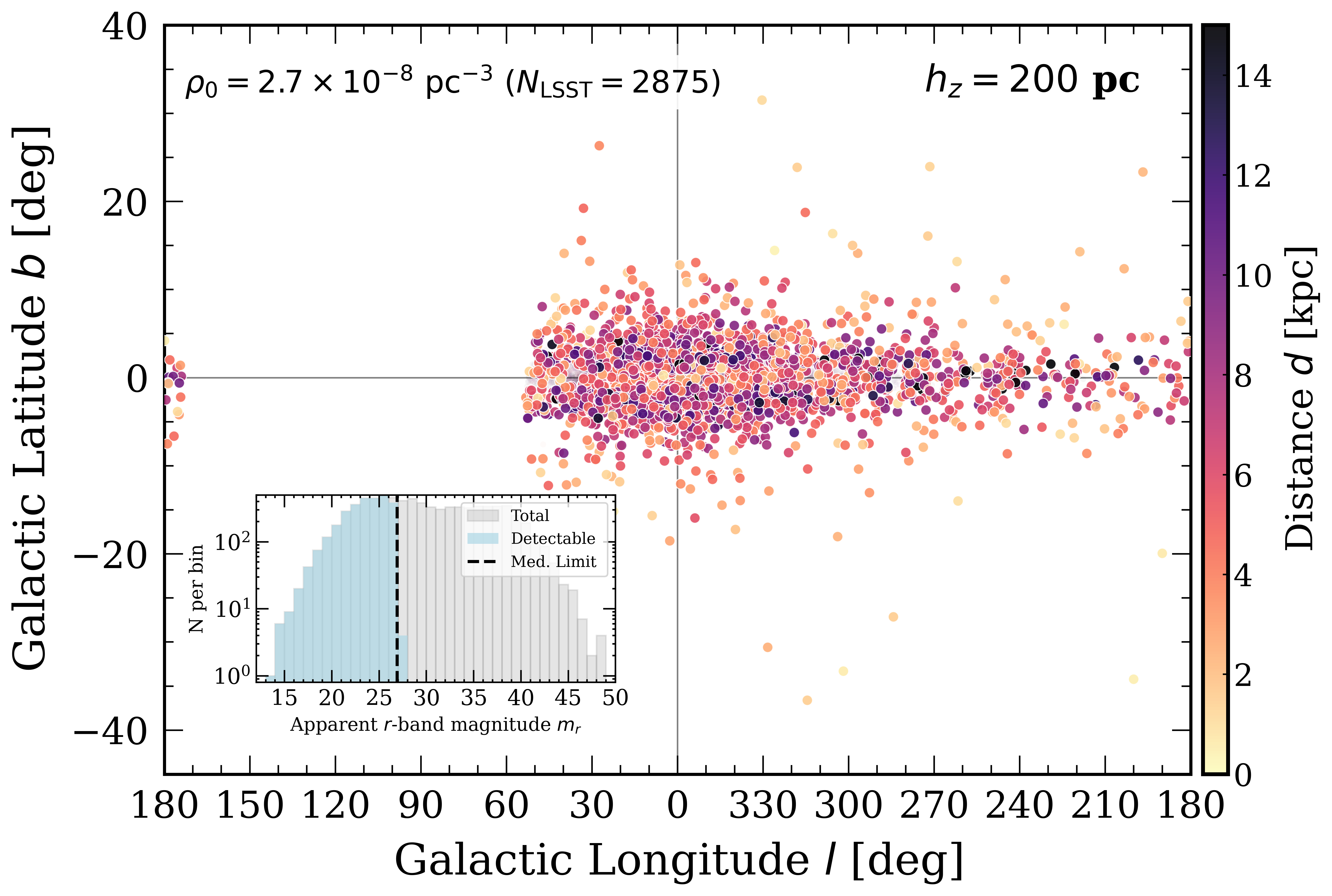} 
    \includegraphics[width=0.49\linewidth]{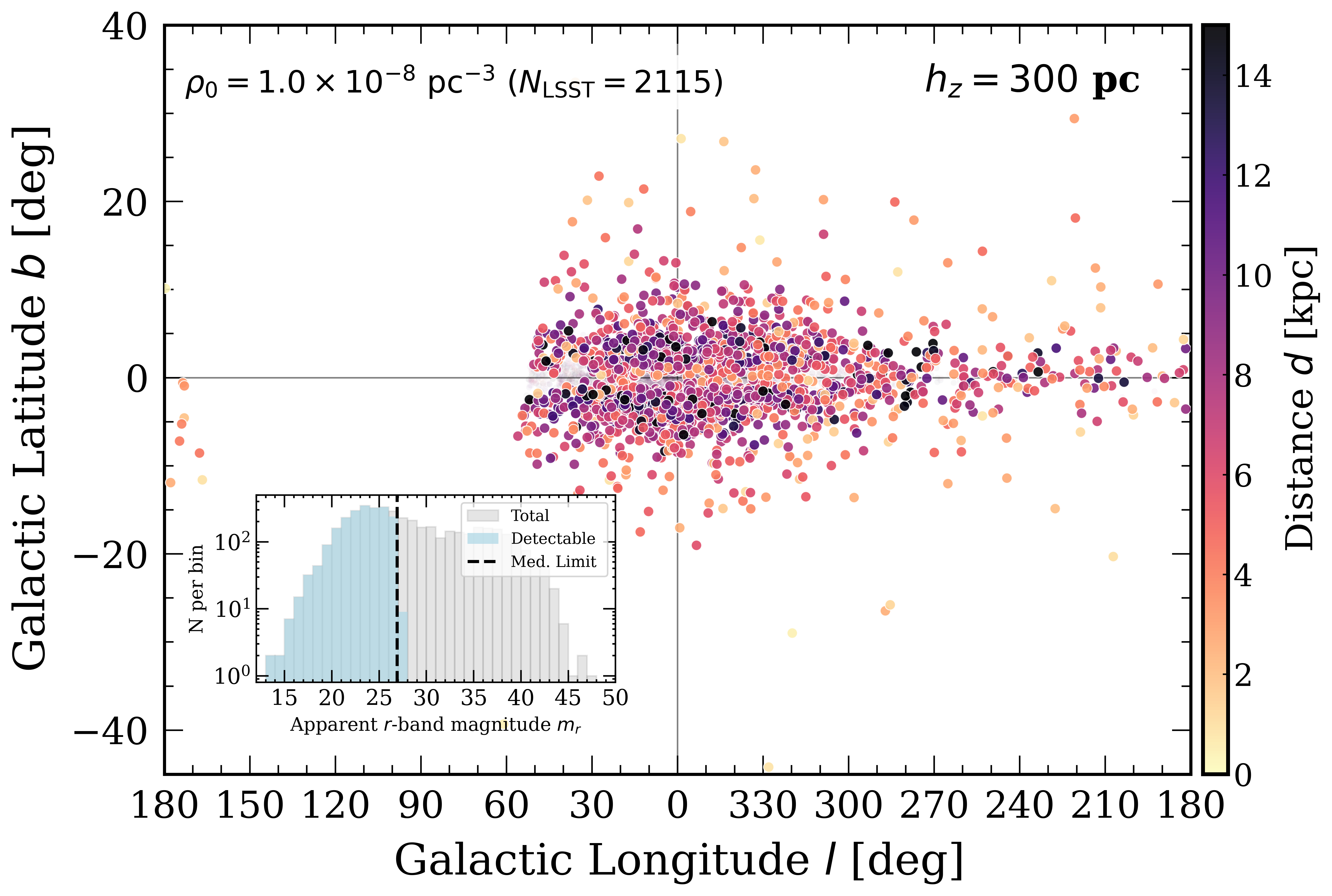}
    \includegraphics[width=0.49\linewidth]{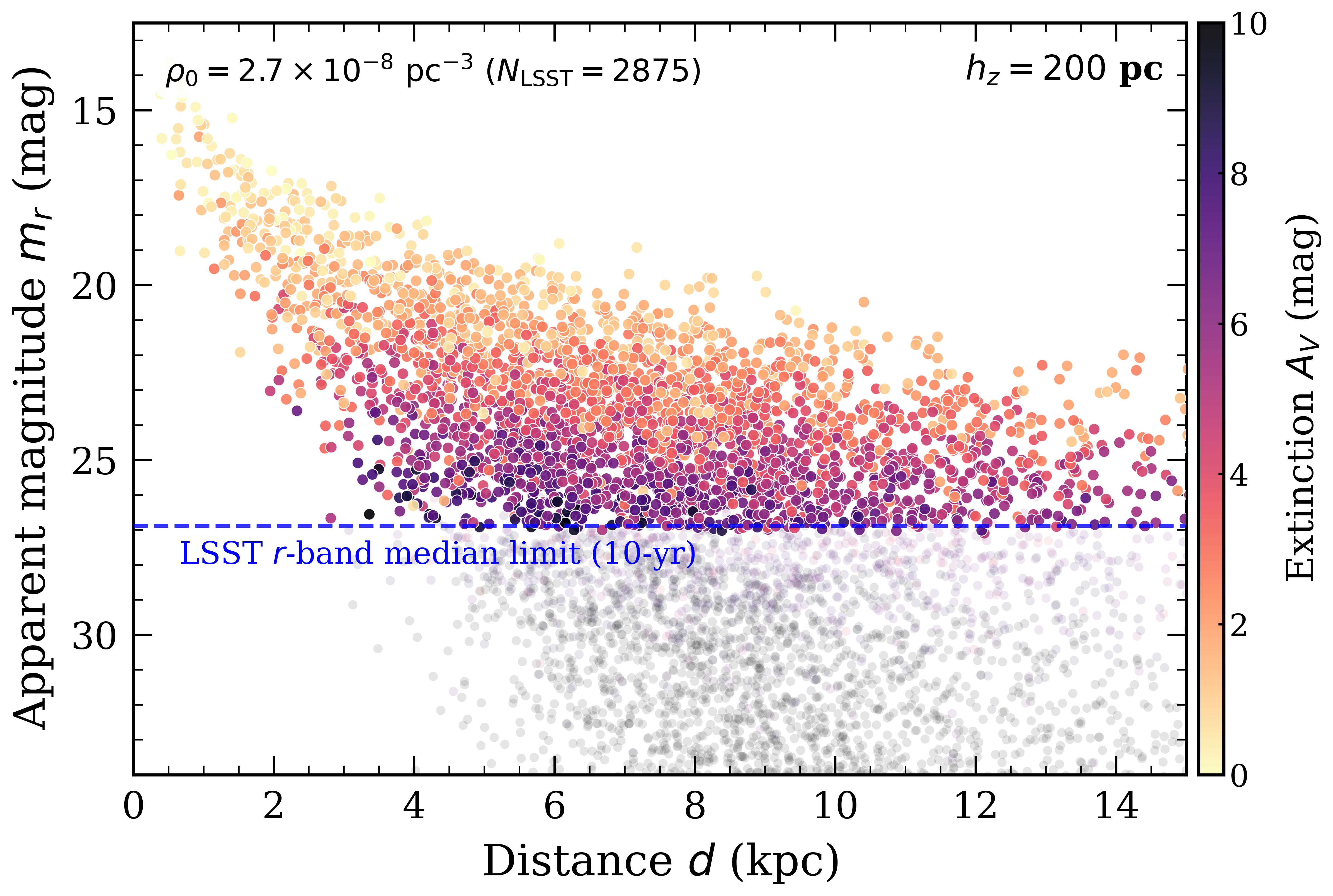} 
    \includegraphics[width=0.49\linewidth]{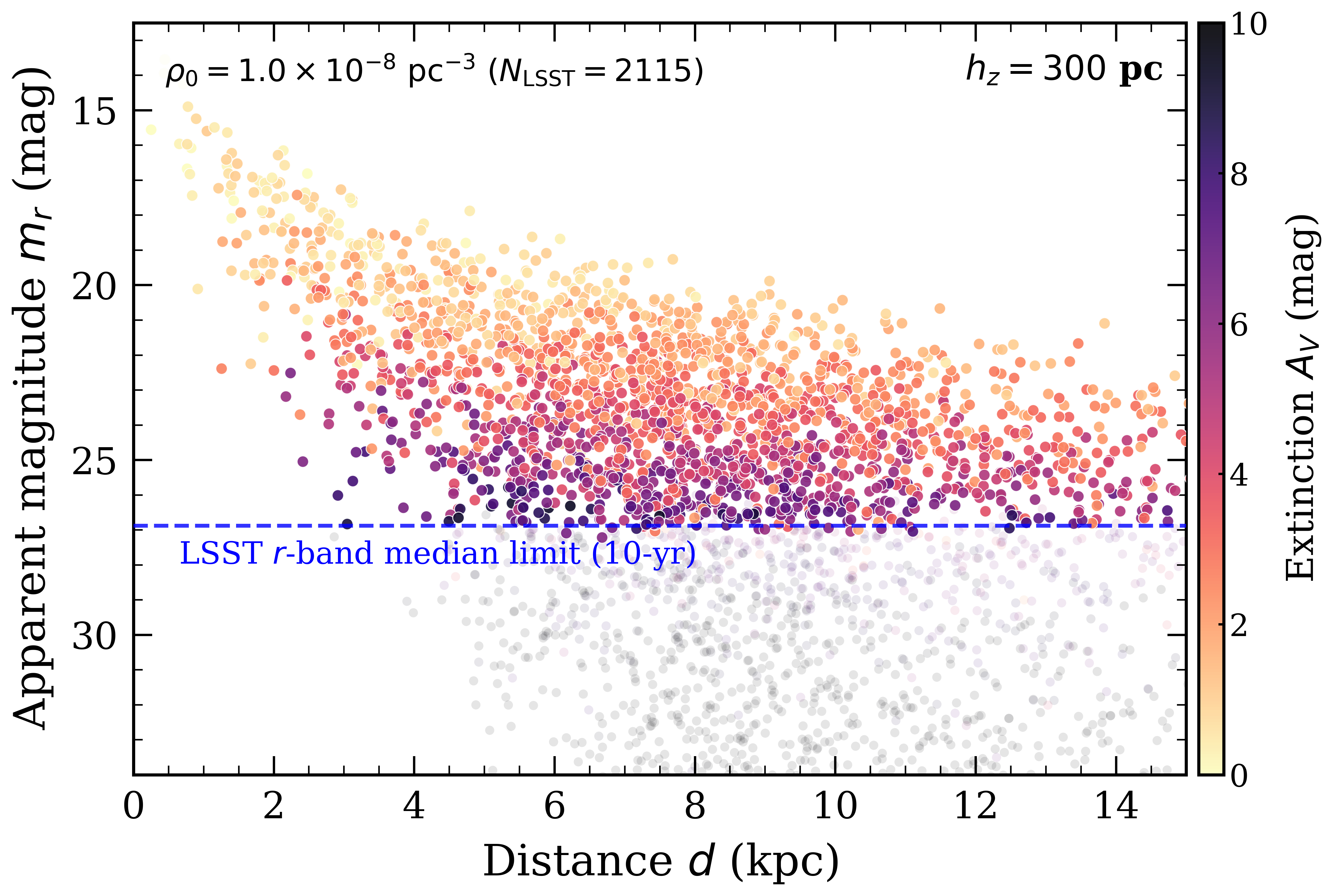}
    \caption{Detectability of the high-state AM CVn population in the Milky Way with LSST  during the 10-year survey. The  number of detectable systems is estimated based on a local space density of $\rho_0=2.7 \times 10^{-8}\ \text{pc}^{-3}$ ($h_z=200$~pc, left panels) and $\rho_0=1.0 \times 10^{-8}\ \text{pc}^{-3}$ ($h_z=300$~pc, right panels) for a single realization. The color bars indicate the distance to the targets (top) and the Galactic extinction $A_V$ at the source distance (bottom).  Top panel: Distribution of objects in Galactic coordinates. The sub-panel shows the distribution of objects over the apparent $r-$filter magnitude. The gap at Galactic longitude $l\approx 60^{\circ}-170^{\circ}$ is caused by the LSST survey strategy, leading to omission of about 9\% of the total population. Bottom panel: Apparent $r-$filter magnitude versus the distance. The horizontal line shows the median $r-$filter magnitude for the LSST 10-year survey ($m_{r}\lesssim26.9$~mag). About $34-44$\% of the total population is expected to be detectable by LSST during the survey at a median limit of $m_{r}\lesssim26.9$~mag.} 
    \label{fig:LSST}
\end{figure*}

\subsubsection{Detectability with next-generation observatories: LISA, TianQin and LSST}
\label{sec:detectability}

Using our empirically estimated space density of high-state AM CVn binaries, we calculated the total number of objects detectable by future millihertz GW observatories (LISA and TianQin) over a 4-year observation period and by the 10-year LSST  survey. 

We used the total number of high-state AM CVns in the Milky Way (see Table~\ref{tab:amcvn_detections}), as computed from Eq.~(\ref{eq:ntotal}). We assumed that the spatial distribution of objects follows a double exponential disk, according to Eq.~(\ref{eq:rho_model}). We sampled their locations in the Milky Way within the following limits: $R \in [0, 16]$~kpc, $z \in [-5h_z, 5h_z]$, and $\phi \in [0, 2\pi]$. The heliocentric distance to the objects was computed in the cylindrical coordinate system, assuming a solar distance of $R_{Sun}$:
\begin{gather}
    d =\sqrt{R^2 + R_{\rm Sun}^2 - 2 R R_{\rm Sun} \cos \phi + z^2}.
    \label{eq:coord1}
\end{gather}

We assumed a uniform distribution for the orbital periods of high-state AM CVn binaries within the $5-30$ minute range. This wide period range is motivated by several known high-state systems observed with periods between 20 and 30 minutes (see Table~\ref{tab:known}). The WD mass was sampled from a Gaussian distribution with a mean and variance of $0.83 \pm 0.23 \, M_{\odot}$ \citep{2011A&A...536A..42Z}\myfootref{note:wdmass}. To sample the donor mass, we used evolutionary track where the mass ratio follows the orbital period (see Section~\ref{sec:donor} and the right panel of Fig.~\ref{fig:J2007_fit}). The evolutionary tracks for the WD and evolved CV channels show similar functional forms for the mass ratio, $q \propto P_{\rm orb}^{-1.37}$, whereas the helium star channel follows a distinct evolutionary path and does not produce systems with orbital periods below 10~minutes. We adopted the WD channel's  $q \propto P_{\rm orb}^{-1.37}$ as the representative function for our simulations. We assumed that across the $5-30$~minute orbital period range (starting from the period minimum and toward longer periods), the mass ratio varies from 0.22 down to 0.02. For each simulated target, we computed the characteristic strain for a 4-year observation period (see Section~\ref{sec:gw}). We assumed that a system is detected by a GW mission if its signal-to-noise ratio is $(S/N) \ge 5$. In Appendix~\ref{app:population}, we present simulations for the helium star donor channel to investigate the impact of the formation channel on our results.

To estimate the total number of detectable systems by the LSST during 10-year survey, we assumed that the accretion disk is the primary source of emission in the optical band for all simulated high-state AM CVns with orbital periods of $5-30$ minutes. We sampled the accretion rates for systems in the high state from a uniform distribution in the $log\dot{M} \in [-9, -7]$ range. The outer radius of the accretion disk was computed using Eq.~(\ref{eq:R_out}), while the inner radius was assumed to be equal to the WD radius. The inclination angle was sampled from $sin(i)$ distribution. We sampled the locations of objects in the Mike Way and calculated the  Galactic coordinates ($l,b$) as: 
\begin{gather}
    \sin b = \frac{z}{d}\quad  \text{and}\quad   l = \operatorname{atan2}(R \sin \phi, R_{\rm Sun} - R \cos \phi).
    \label{eq:coord2}
\end{gather}

Using these coordinates and the distances to the objects, we estimated the Galactic extinctions $A_V$ from the {\tt Combined19map}. Based on the distances, Galactic extinctions  and the accretion disk's SED (see Eqs.~(\ref{eq:tdisk}) and (\ref{eq:fdisk})), we computed the apparent $r$-filter magnitude of the simulated  population. We used the Rubin Observatory's baseline\footnote{\url{https://survey-strategy.lsst.io/baseline/index.html}} survey map (v.5.1.0) to estimate the 10-year $r-$filter limiting magnitude at each object's Galactic coordinates. Due to the Rubin Observatory's  observing strategy, about 9\% of the systems will be missed during the 10-year survey.

Fig.~\ref{fig:GWpopulation} shows the characteristic strain versus the GW frequency for a single realization of the high-state AM CVns population, assuming  disk scale heights of $h_z=200$~pc and $h_z=300$~pc. The current sample of verification and detectable AM CVn systems appears as a set of outliers relative to the predicted Galactic population of high-state systems. The majority of the known systems are located at distances less than 2 kpc. Our simulations suggest that the bulk of the population exists at greater distances $d\gtrsim$ 2 kpc, concentrating towards the Galactic center and inner disk region. This indicates a selection bias in electromagnetic surveys towards detecting the nearest sources in the Solar neighborhood, missing the intrinsically faint population at larger distances. Simulations of the high-state AM CVn population based on the helium star donor channel show better agreement with the known detectable and verification AM CVn systems at long orbital periods, potentially explaining part of the observed discrepancy (see Appendix~\ref{app:population} and Fig.~\ref{fig:GWpopulation_app}). This alignment is likely driven by the higher chirp masses and elevated mass accretion rates of high-state AM~CVn systems evolving via the helium star channel. However, these simulations do not account for the short-period systems ($P_{orb}\lesssim 10$ minutes) and do not completely rule out a selection bias in electromagnetic surveys toward detecting the nearest sources. 

Fig.~\ref{fig:LSST} shows the distribution of the simulated population in Galactic coordinates (top panel) and the apparent $r-$filter magnitude versus distance (bottom panel). Systems with low extinction ($A_V \approx 0-2$ mag) can be detectable up to 15 kpc, while those with high extinction ($A_V\gtrsim6$ mag) fall below the 10-year LSST detection limit.

We performed $10\ 000$ Monte Carlo iterations, sampling the locations and parameters of the objects, to estimate the median number of detectable binaries with LISA, TianQin and LSST (see Table~\ref{tab:amcvn_detections}). We estimated that LISA will detect  about $N_{\rm LISA} \approx 1410-4810$ ($\approx 850-2580$) systems, assuming a Galactic disk model with $h_z = 200$~pc ($h_z = 300$~pc). TianQin  is expected to detect  about $N_{\rm TianQin}\approx 760-2585$ ($\approx 460-1380$) systems, while LSST will detect about $N_{\rm LSST}\approx 1350-4610$ ($\approx 1020-3115$) systems, assuming a similar Galactic disk model with $h_z = 200$~pc ($h_z = 300$~pc). Our simulations indicate that approximately $\approx36\%$ of the total high-state AM~CVn population in the Milky Way is expected to be detectable by LISA and $\approx19 \%$ by TianQin during a 4-year mission with a signal-to-noise ratio $(S/N) \ge 5$, while $\approx34-44\%$ should be detectable by the LSST during 10-year survey at a median limit of $m_{r}\lesssim26.9$~mag.

We note that we ran our simulations assuming uniform distributions for both the orbital periods and mass accretion rates of high-state AM CVn binaries, as their true intrinsic  distributions remain uncertain. To investigate how alternative distributions impact our simulation results, we used the evolutionary models from Fig.~\ref{fig:models}. For the WD channel, the mass accretion rate scales with time since the onset of mass transfer as $\dot{M} \propto t^{-1.30}$ and with the orbital period as $\dot{M} \propto P_{\rm orb}^{-4.67}$, which yields a period evolution of $P_{\rm orb} \propto t^{0.28}$. The high-state phase, defined where $\dot{M}\gtrsim\dot{M}_\mathrm{crit}^{+}$ (see Fig.~\ref{fig:models} and Eq.~(\ref{eq:Mdot_crit})), corresponds only to a period range of $6-18$ minutes. Running the simulation with these distributions increases the number of detectable sources by a factor of $\sim 1.1$ for LISA, but decreases it by a factor of $\sim 1.7$ for TianQin and $\sim 1.1$ for LSST (see Appendix~\ref{app:population} for the helium star channel). More accurate evolutionary calculations are required to robustly constrain the intrinsic period and accretion-rate distributions, which currently remain model-dependent. While a detailed study of these mechanisms is beyond the scope of this paper, upcoming facilities will help constrain these population properties. Future GW mission like LISA and TianQin will be able to observe this hidden population, providing a more complete census of the binaries in the Milky Way.

\section{Conclusion and Summary}
\label{sec:summary}

We present a systematic search for high-state AM CVn binaries in the Milky Way using Gaia and ZTF data. We first selected candidate systems  based on a Gaia color-magnitude cut above the WD sequence (see Fig.~\ref{fig:hr}). We then performed ZTF period search  analyses of these candidate systems, focusing on variable targets with periods in the $5-30$ minute range. We conducted the spectroscopic and photometric follow-up observations with Keck~I/LRIS and CHIMERA, respectively, to confirm the nature of the new targets. We discovered three new variable systems and recovered three previously known objects (see Table~\ref{tab:known} in Appendix~\ref{app:list}). These six systems were used to constrain the local space density and birth rate of the high-state AM~CVn population in the Milky Way. Our main results are summarized as follows:

\begin{enumerate}
\setlength{\itemsep}{0.5em}

    \item Three new high-state AM CVn binaries, ZTF~J1840$-$1742, ZTF~J2007$-$0527 and ZTF~J2111+3158, show orbital periods of 16.86, 18.63 and 16.71 minutes, respectively. These photometric periods are consistent with spectroscopic periods (see Figs.~\ref{fig:chimera} and \ref{fig:rvs}). All three systems are expected to be detectable sources for future GW missions (LISA and TianQin) during a 4-year mission (see Fig.~\ref{fig:gw}).

    \item  The optical spectra of the three new systems show strong helium emission lines with the absence of detectable hydrogen (see Fig.~\ref{fig:spectra}). The helium emission lines are double-peaked in ZTF~J1840$-$1742 and ZTF~J2111+3158, where Doppler tomograms reveal clear accretion disk structures. The helium emission lines of  ZTF~J2007$-$0527 are single-peaked; the corresponding Doppler tomogram shows an extended feature resembling a centrally-filled disk (see Fig.~\ref{fig:doppler}). These three objects expand the limited sample of high-state AM CVns exhibiting observable accretion disks (see Table~\ref{tab:known}).

    \item We estimated the $3\sigma$ X-ray luminosity upper limits to be on the order of $\rm 10^{33}\ erg\ s^{-1}$.  Based on SED modeling, we estimated the accretion rate of these systems on the order of  $10^{-9}\,M_{\odot }\,\text{yr}^{-1}$ (see Table~\ref{tab:params}). These accretion rates are consistent with those typical for AM CVns in the high-state (see Fig.~\ref{fig:mdot}).

    \item We inferred the local space density of the high-state AM~CVn population in the Milky Way to be  $\rho_0 = (2.7 \pm 1.5) \times 10^{-8}~\text{pc}^{-3}$ ($h_z = 200$~pc) and $\rho_0 = (1.0\pm 0.5) \times 10^{-8}~\text{pc}^{-3}$ ($h_z = 300$~pc) (see Table~\ref{tab:amcvn_detections}). These systems  represent only $\approx 2-5\%$ of the total AM CVn population density in the Milky Way \citep[$\sim 6 \times 10^{-7}~\text{pc}^{-3}$;][]{2013MNRAS.429.2143C,2022MNRAS.512.5440V,2025PASP..137a4201R}. 
    
    \item The birth rate of high-state AM~CVns, $R_{\text{birth}}\approx(4.3\pm2.4)\times 10^{-4}~\text{yr}^{-1}$(for $h_z = 200$~pc) and $R_{\text{birth}}\approx(2.4\pm1.2)\times 10^{-4}~\text{yr}^{-1}$ (for $h_z = 300$~pc), is statistically consistent with that of the total Galactic AM~CVn population (see Section~\ref{sec:population}). This implies that most systems entering the high-state survive and evolve toward the long-period low-state ($P_{orb}\gtrsim 30\,$ minutes). However, due to the current limited sample of AM~CVns, we cannot rule out that some systems might fail to survive the high-state phase, likely resulting in stellar mergers or helium-triggered thermonuclear explosions, such as Type~.Ia supernovae.

    \item  Our empirical space density of high-state AM CVn population suggests that between  2350 and 13330 systems exist in the Milky Way. Future GW observatories, LISA and TianQin, are expected to detect about $\approx36\%$  and  $\approx19\%$ of the total high-state AM~CVn population, respectively, during a 4-year mission with a signal-to-noise ratio $(S/N) \ge 5$. The Rubin Observatory's Legacy Survey of Space and Time (LSST) will detect about  $\approx34-44\%$  of the systems during its 10-year survey at a median limit of $m_{r}\lesssim26.9$~mag (see Table~\ref{tab:amcvn_detections} and Figs.~\ref{fig:GWpopulation} and~\ref{fig:LSST}).

\end{enumerate}

Through a systematic and targeted search, our work has placed strong constraints on the Galactic space density and total population of ultra-compact AM CVn binaries in a high accretion state. The bright nature of the sources presented here ($G\sim18$ mag) provides an exciting opportunity for future LSST discoveries of more distant, fainter AM CVns. Most importantly, our work has established a baseline for detectability in GWs by upcoming missions such as LISA and TianQin; the GW signals detected by those missions will provide independent constraints on space density as well as on stellar parameters, informing our understanding of the binary evolution of these objects. In conclusion, this survey represents an important step in multi-messenger astrophysics, establishing a baseline space density using electromagnetic surveys that will inform the GW surveys of decades to come.

\begin{acknowledgements}

Based on observations obtained with the Samuel Oschin Telescope 48-inch and the 60-inch Telescope at the Palomar Observatory as part of the Zwicky Transient Facility project. ZTF is supported by the National Science Foundation under Grants No. AST-1440341 and AST-2034437 and a collaboration including current partners Caltech, IPAC, the Weizmann Institute of Science, the Oskar Klein Center at Stockholm University, the University of Maryland, Deutsches Elektronen-Synchrotron and Humboldt University, the TANGO Consortium of Taiwan, the University of Wisconsin at Milwaukee, Trinity College Dublin, Lawrence Livermore National Laboratories, IN2P3, University of Warwick, Ruhr University Bochum, Northwestern University and former partners the University of Washington, Los Alamos National Laboratories, and Lawrence Berkeley National Laboratories. Operations are conducted by COO, IPAC, and UW. This work has made use of data from the European Space Agency (ESA) mission Gaia (https://www.cosmos.esa.int/gaia), processed by the Gaia Data Processing and Analysis Consortium (DPAC, https://www.cosmos.esa.int/web/gaia/dpac/consortium). Funding for the DPAC has been provided by national institutions, in particular the institutions participating in the Gaia Multilateral Agreement. Some of the data presented herein were obtained at Keck Observatory, which is a private 501(c)3 non-profit organization operated as a scientific partnership among the California Institute of Technology, the University of California, and the National Aeronautics and Space Administration. The Observatory was made possible by the generous financial support of the W. M. Keck Foundation. We wish to recognize and acknowledge the very significant cultural role and reverence that the summit of Maunakea has always had within the Native Hawaiian community. We are most fortunate to have the opportunity to conduct observations from this mountain. We are grateful to the staffs of the Palomar and Keck Observatories for their work in helping us carry out our observations. 

I.G. work was funded by a grant from the Academy of Sciences of the Republic of Tatarstan provided to higher education institutions, scientific and other organizations to support human resource development plans in terms of encouraging their research and academic staff to defend doctoral dissertations and conduct research activities (Agreement No. 12/2025-PD-KFU dated December 22, 2025). V.D. and A.S. acknowledge support from Kazan Federal University. A.C.R acknowledges support from an NSF Graduate Fellowship, a Future Faculty Leader Fellowship, and the Institute for Theory and Computation at the Center for Astrophysics $|$ Harvard \& Smithsonian. We thank the anonymous referee for the insightful feedback and comments, which helped improve this manuscript.
\end{acknowledgements}

\bibliographystyle{aa} 
\bibliography{highstateamcvn}

\begin{appendix}
\nolinenumbers

\section{ List of known high-state and direct-impact AM CVn binaries} 
\label{app:list}

\begin{table*}
\caption{ List of known high-state and direct-impact AM CVn binaries.
	\label{tab:known}}
\renewcommand\arraystretch{1}
\setlength\tabcolsep{3.pt}
\centering
\begin{tabular}{l c c c c c c c c c}
\hline
\hline
Name & R.A. & Decl. & $P_{orb}$ & Gaia ID & $G$ & $\varpi$ & $\sigma_{\varpi}$ & System Type &  Discovery  \\ 
 & (deg) & (deg) & (min) & & (mag) & (mas) & (mas) & & References \\ 
\hline

ES Cet$^{\star}$ & 30.21772 & $-$9.4088 & 10.3 & 2462596293177188480 & 16.8 & 0.561 & 0.068 & High-state & (1) \\

SDSS J1351$-$0643$^{\star}$ & 207.9769 & $-$6.7192 & 15.7 & 3620363418542428416 & 18.7 & 0.658 & 0.220 & High-state & (2) \\

ZTF J2111+3158$^{\star\star}$ & 317.8310 & 31.9795 & 16.7  & 1853123945507031808 & 18.0 & 0.808 & 0.097 & High-state & This work \\

ZTF J1840$-$1742$^{\star\star}$ & 280.2187  & $-$17.7136  & 16.9  &4099443479729561472 & 18.1 & 0.744  & 0.147 & High-state & This work \\

AM CVn & 188.7278 & 37.6290 & 17.1 & 1519860699806445184 & 14.1 & 3.311 & 0.030 & High-state & (3) \\

SDSS J1908+3940 & 287.0711 & 39.6768 & 18.2 & 2100480767163455360 & 16.2 & 1.023 & 0.034 & High-state & (4) \\

HP Lib$^{\star}$ & 233.9710 & $-$14.2201 & 18.4 & 6265476408553544320 & 13.6 & 3.567 & 0.031 & High-state & (5) \\

ZTF J2007$-$0527$^{\star\star}$ & 301.8123  & $-$5.4645 & 18.6  & 4220524281429040128  & 17.7 & 0.822 & 0.107 & High-state & This work \\

CXOGBS J1751$-$2940$^\dagger$ & 267.7819 & $-$29.6771 & 22.9 & 4056461714848807296 & 17.5 & 1.061 & 0.108 & High-state & (6) \\

TIC 378898110$^\dagger$ & 180.9110 & $-$60.3801 & 23.0 & 6058834949182961536 & 14.3 & 3.233 & 0.019 & High-state & (7) \\

\hline

HM Cnc & 121.5956 & 15.4586 & 5.35 & 654879021108862464 & 20.9 & -- & -- & Direct-impact accretor  & (8) \\

eRASSU J0608$-$7040 & 92.1646 & $-$70.6706 & 6.20 & -- & -- & -- & -- & Direct-impact accretor & (9) \\

ZTF J0546+3843 & 86.6143 & 38.7204 & 7.95 & 189913111550827264 & 19.3 & 0.086 & 0.309 & High-state & (10) \\

ATLAS J1013$-$4516 &153.4270  &$-$45.2824  & 8.56 & 5414242958825679488  & 19.5  & 0.199  & 0.280 & Direct-impact accretor & (11) \\

ZTF J1858$-$2024 & 284.5248 & $-$20.4135 & 8.68 & 4085034074899981312 & 19.4 & 0.754 & 0.399 & High-state & (10) \\

V407 Vul & 288.6087 & 24.9454 & 9.48 & 2023676031667286016 & 19.4 & 0.098 & 0.238 & Direct-impact accretor  & (12) \\

ZTF J0425+3858 & 66.4592 & 2.5988 & 13.2 & -- & -- & -- & -- & High-state & (10) \\

ZTF J1905+3134 & 286.2972 & 31.5757 & 17.2 & 2042772624503621632 & 20.7 & 1.865 & 1.543 & High-state & (13) \\

3XMM J0510$-$6703 & 77.6444 & $-$67.0651 & 23.6 & 4661852408981634688 & 20.9 & -- & -- & Direct-impact accretor  & (14) \\

ZTF J2228+4949 & 337.1128 & 49.8212 & 28.6 & 1988156068027106304 & 19.2 & 0.155 & 0.222 & High-state & (13) \\
\hline

\end{tabular}
\flushleft
{\bf Notes.} ($^{\star}$) Known systems recovered in this study. ($^{\star\star}$) New objects discovered in this study. ($^\dagger$) Target declination is too far south for the ZTF survey footprint. \\
{\bf References:} (1)  \citet{2002PASP..114..129W}; (2) \cite{2018MNRAS.477.5646G}; (3) \cite{1967AcA....17..255S}; (4) \cite{2011ApJ...726...92F}; (5) \cite{1994MNRAS.271..910O}; (6) \cite{2016MNRAS.462L.106W}; (7) \cite{2024MNRAS.527.3445G}; (8) \cite{1999A&A...349L...1I}; (9) \cite{2024A&A...683A..21M}; (10) \cite{2024ApJ...977..262C}; (11)  \cite{2026arXiv260107925C}; (12) \cite{1996A&A...307..459M}; (13) \cite{2020ApJ...905...32B}; (14) \cite{2017A&A...598A..69H}.

\end{table*}

We compiled a list of known high-state AM CVn binaries (see Table \ref{tab:known}). The majority of confirmed objects were adopted from the \citet{2025A&A...700A.107G} catalog.  Gaia source identifiers (Gaia IDs), magnitudes ($G$), and parallax ($\varpi$)  and its uncertainty ($\sigma_{\varpi}$) were adopted from Gaia DR3. 

In our target selection (see Section~\ref{sec:survey} and Fig.~\ref{fig:hr}), we identified ten high-state AM CVns with a parallax signal-to-noise ratio of $\varpi/\sigma_{\varpi} \ge 2$. Using our period search analyses, we recovered three known objects (indicated with a single asterisk, $^{\star}$) and discovered three new objects (indicated with a double asterisk, $^{\star\star}$), resulting in a final sample of six objects. The systems CXOGBS J1751$-$2940 and TIC 378898110 were excluded because they did not pass our ZTF declination criteria ($\delta \geq -28^{\circ}$). We did not detect periodic signals within our period search analyses for AM CVn and SDSS J1908+3940. The remaining known objects from the catalog have $\varpi/\sigma_{\varpi} < 2$ and were consequently excluded from our analysis. 

The object ASASSN$-$14cc (marked in Fig.~\ref{fig:mdot}) was excluded from our sample because it is an outbursting system rather than one with a persistently stable accretion disk. We also exclude SDSS~J1831+4202 \citep{2023MNRAS.524.4867I} from our analysis since it is not securely listed as a high-state system in the collection of systems by \cite{2025A&A...700A.107G}, and features short-term and long-term outbursts that may exclude it from the list of high-state systems altogether.

\section{Light curve modeling of ZTF~J2007$-$0527}
\label{app:LC}

\begin{figure*}
    \centering
    \includegraphics[width=1\linewidth]{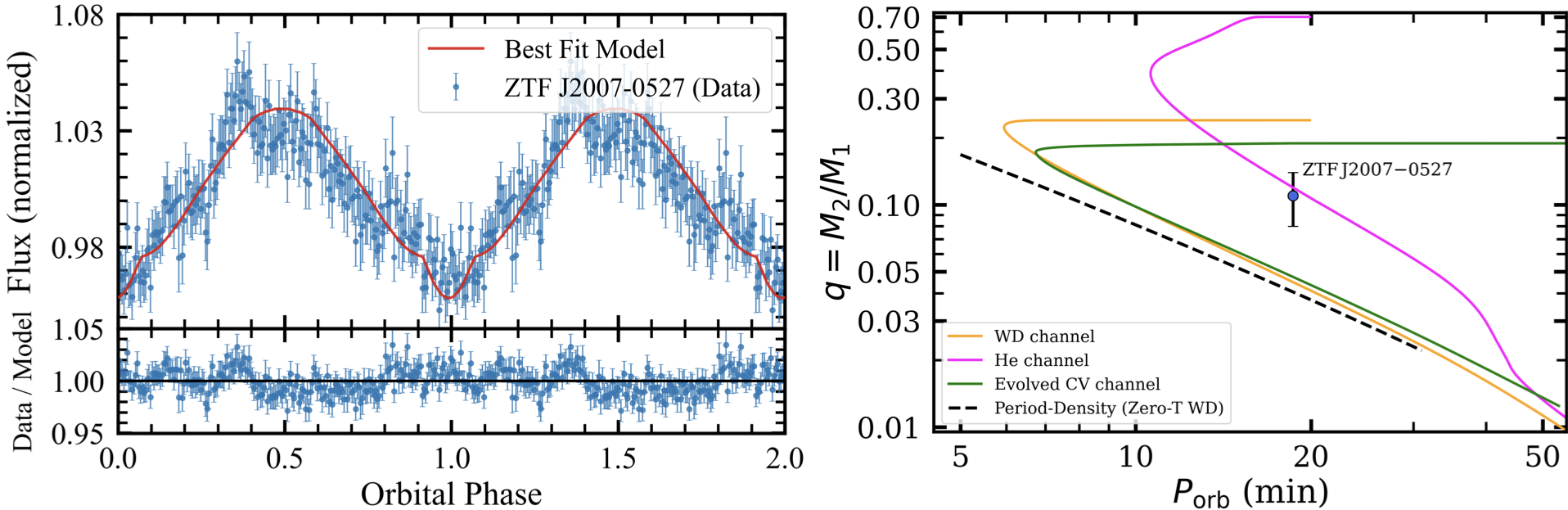}
    \caption{Left: Phase-folded  CHIMERA light curve of ZTF~J2007$-$0527 (blue points) along  with  the best-fit model (red line). Right: Mass ratio versus the orbital period for the three evolutionary channels of AM CVn systems: the WD channel (orange; \citealt{2021ApJ...923..125W}), the helium star channel (magenta; \citealt{2015ApJ...807...74B, 2021ApJ...923..125W}), and the evolved CV channel (green; \citealt{2023A&A...678A..34B}). The black dashed line shows the relation computed from the period-density, using a WD mass of $\rm M_{\rm WD}\approx0.8\,M_\odot$ and zero-temperature WD mass-radius \citep{1988ApJ...332..193V} relations for the donor. The measured mass ratio of ZTF~J2007$-$0527 is most consistent with formation via the helium star channel, rather than the WD or evolved CV channels.}
    \label{fig:J2007_fit}
\end{figure*}

While we attempted to model light curves of all targets, the analyses for ZTF~J1840$-$1742 and ZTF~J2111+3158 yielded unphysical binary parameters. This likely reflects the complex non-LTE effects or non-steady-state accretion disk structures in these systems. We focus our light curve analysis on ZTF~J2007$-$0527, where the blackbody approximation provided physically consistent parameters.

We modeled the CHIMERA light curves of ZTF~J2007$-$0527 using a three-component model: a spherically symmetric WD, a Roche-lobe filling donor star, and an axisymmetric accretion disk of constant vertical thickness. We fixed the WD and donor temperatures at conservative values of $T_{\rm WD} = 16000\,{\rm K}$ and $T_\mathrm{donor} = 3000\,{\rm K}$. We experimented with different donor temperatures in the $\rm 2000-5000$ K range, and the WD temperature within  $\rm 10 000-20 000$ K, finding that these variations do not affect the final results. A bright spot component was not included in the model, as its inclusion does not affect the light curve modeling. 

The surfaces of all components (WD, donor and accretion disk) are discretized into finite elements, each characterized by its centroid coordinates, surface normal vector, area, and effective temperature. The contribution from each surface element to the total observed flux is computed by integrating the element's blackbody spectrum over the filter transmission curve. The accretion disk is assumed to be in steady state, and its temperature profile follows the standard prescription of \cite{1973A&A....24..337S}. We included the linear limb darkening of the WD and gravity darkening of the donor. We assumed that the donor surface is irradiated by X-rays  originating in the boundary layer at the surface of the WD. The effective temperature of the irradiated donor is determined by the energy balance equation:

\begin{equation}
    \sigma T_{\mathrm{eff}}^4 = \sigma T_{\mathrm{eff,0}}^4 + \kappa f_X,
\end{equation}

\noindent where $T_{\rm eff,0}$ is the unirradiated effective temperature, $\kappa=0.4$ is the X-ray thermalization coefficient, $f_X$ is the X-ray flux incident at the surface element, and $\sigma$ is the Stefan–Boltzmann constant. We fitted phase-folded light curves using the Goodman--Weare Markov Chain Monte Carlo (MCMC) sampler \citep{2010CAMCS...5...65G} implemented in the \texttt{emcee} Python package \citep{2013PASP..125..306F}. The data were binned into 200 phase bins to improve the signal-to-noise ratio. The orbital period, outer disk radius, and X-ray luminosity were fixed to the values presented in Table \ref{tab:params}. Only inclination angle, mass ratio and WD mass were set as free parameters. 

Fig.~\ref{fig:J2007_fit} (left panel) shows the modeled CHIMERA light curve for ZTF~J2007$-$0527. The model resulted in a reduced chi-square of $\chi^2/\mathrm{dof}= 1.18$. We estimated the mass ratio, inclination angle and WD mass of ZTF~J2007$-$0527 to be $q=0.11^{+0.03}_{-0.02}$, $i =69^{\circ}\pm2$ and $M_{\rm WD}=0.61^{+0.14}_{-0.06}\ M_{\odot}$, respectively. The errors correspond to a 68\% confidence level. The irradiated surface of the donor star, partially eclipsed by the accretion disk, is the  dominant contribution to the the variability in observed light curve of ZTF~J2007$-$0527. Irradiation raises the characteristic temperature of the heated face to $\approx 4900$~K. Such irradiation effects are consistent with recent studies emphasizing the role of donor irradiation in CVs \citep{2026arXiv260515897D}, including detections of irradiated donors in long-period AM CVn systems with JWST \citep{2026OJAp....957856E}. The estimated mass ratio is consistent with the value expected for the helium star channel, rather than WD or evolved CV channels (see Fig.~\ref{fig:J2007_fit}, right panel). We additionally inspected WD and evolved CV models with different initial parameters. The evolved-CV channel can produce donors as bloated as those in the helium-star channel, but for ZTF~J2007$-$0527 this would correspond to the pre-period-minimum stage, when the system is still evolving toward shorter orbital periods. The evolved CV models reach accretion rates of $\dot{M}\approx 10^{-9}\,M_\odot\,\mathrm{yr}^{-1}$ at $P_\mathrm{orb}=18.63\,$minutes, which is below the rate estimated for ZTF~J2007$-$0527 (see Table~\ref{tab:params}). We therefore consider the helium-star channel to remain the favored interpretation. The current model could not fully reproduce the feature in the observed light curve around orbital phase $\approx 0.3$, likely due to the limitations of the blackbody assumption for the irradiated donor surface and the accretion disk structure.

\section{ ZTF detection probability for high-amplitude high-state AM CVn binaries}
\label{app:ztfrecovey}

We searched for high-state AM CVn binaries showing periodic signals in the $5-30$ minute range. Our ZTF period search analysis is sensitive to various types of variability, such as partial eclipses, ellipsoidal modulation of the donor, bright spots, and disk asymmetries.  The observed amplitude ($A$) of the variability depends on the inclination angle of the system, $A(i) = A_{90} \sin i$, where $A_{90}$ is the intrinsic edge-on ($i=90^\circ$) amplitude. If the observed amplitude falls below the detection threshold, we do not expect to detect the signal. The recovery fraction of the signal depends on the observed amplitude  and the number of epochs ($N$) in the ZTF light curve, $f_{\rm rec}(A(i), N)$. We assumed that the random orientation of the AM CVn binaries follows the $\sin (i)$  inclination distribution. The total detection probability of the periodic signal in the ZTF light curve, $P_{\rm det}$, with amplitude $A_{90}$ is given by:
\begin{gather}
    P_{ \rm det} = \int_{0}^{\pi/2} f_{\rm rec}(A(i), N) \times \sin (i) \, di \ ,
    \label{eq:Pdet}
\end{gather}
where this equation simultaneously accounts for the recovery fraction of the periodic signal in the ZTF light curve, $f_{\rm rec}(A(i), N)$, and the geometric selection effects caused by the random orientation of AM~CVn orbits ($\sin i$).

We generated light curves with controlled variability to estimate the recovery fraction $f_{\rm rec}(A, N)$ for each of the six high-state AM CVn binaries. For each target, we identified non-variable stars within a $10'$ search radius in the ZTF database. Control stars were selected to have a median $r$-filter magnitude close to our targets’ magnitude within a $0.5$ mag difference, and a total number of epochs $N$ within $15\%$ of the targets' epoch in the ZTF light curve. With such selection of the control stars, the noise properties and time sampling of the control ZTF light curves closely match those of our targets. For each control light curve, we injected a sinusoidal periodic signal $S(t)$ defined as:
\begin{gather}
    S(t) = m(t) + A \times \sin\left(\frac{2\pi}{P}t + \varphi_0\right) \ ,
    \label{eq:s}
\end{gather}
where $m(t)$ is the original observed magnitude of the non-variable control star, $P$ is the orbital period of our target, $\varphi_0$ is a random phase chosen from a uniform distribution from 0 to $ 2\pi$, and $A$ is the amplitude in magnitudes. To estimate the $f_{\rm rec}(A, N)$, we used a grid of amplitudes $A=(0.02- 0.35)$ mag and generated 1000 light curves $S(t)$ per amplitude with the period fixed to the estimates for our targets (see Table~\ref{tab:known}). In each realization, we searched for the period using the Lomb-Scargle periodogram and recorded the number of successfully recovered periodic signals. A signal was considered successfully recovered if the period corresponding to the highest peak in the periodogram was within 1\% of the injected period and its significance (defined as the peak power divided by the median absolute deviation of the periodogram) was greater than 30. We then computed the $f_{\mathrm{rec}}(A, N)$ by dividing the number of successfully recovered light curves by the total number of simulated light curves.

The intrinsic edge-on amplitude $A_{90}$ for each object in our sample is not known a priori. The CHIMERA light curves of ZTF~J1840$-$1742, ZTF~J2007$-$0527 and ZTF~J2111+3158 show the observed amplitudes within the 0.9 to 1.1 range in relative flux (see Fig.~\ref{fig:chimera}), corresponding to an intrinsic amplitude of $A_{90} \sim 0.10\ {\rm mag}/\sin(i)$. Light curve modeling of ZTF~J2007$-$0527 gives an inclination angle of $i \approx69^\circ$, corresponding to an intrinsic amplitude of about $A_{90} \sim 0.11$ mag (see Fig.~\ref{fig:J2007_fit}). To account for possible variations of $A_{90}$ among different systems, we computed $P_{\rm det}$ for a range of characteristic values of $A_{90} = (0.05-0.30)$~mag (see Eq.~(\ref{eq:Pdet})). This resulted in detection probabilities for our objects of $P_{\rm det} = (96.5-99.8)\%$. This suggests that for systems having $A_{90}\ga 0.05$ mag, the ZTF detection probability is $\ga 96\%$. The six detected targets in our sample represent the population of high-amplitude variable AM CVns in the high state. However, despite this high detection probability, two other known systems (AM CVn and SDSS J1908+3940) were not detected by our ZTF period search. This implies that our ZTF period search is not sensitive to low-amplitude variables with $A_{90}\la 0.05$ mag, which can be missed entirely with our survey. The completeness of our survey for the total high-state AM CVn population therefore depends on the intrinsic amplitude of $A_{90}$, which we addressed in Section~\ref{sec:population} using empirical constraints from known systems.

\section{ High-state AM CVn population: Impact of the helium star donor channel on detectability}
\label{app:population}

\begin{figure*}
    \centering
    \includegraphics[width=0.49\linewidth]{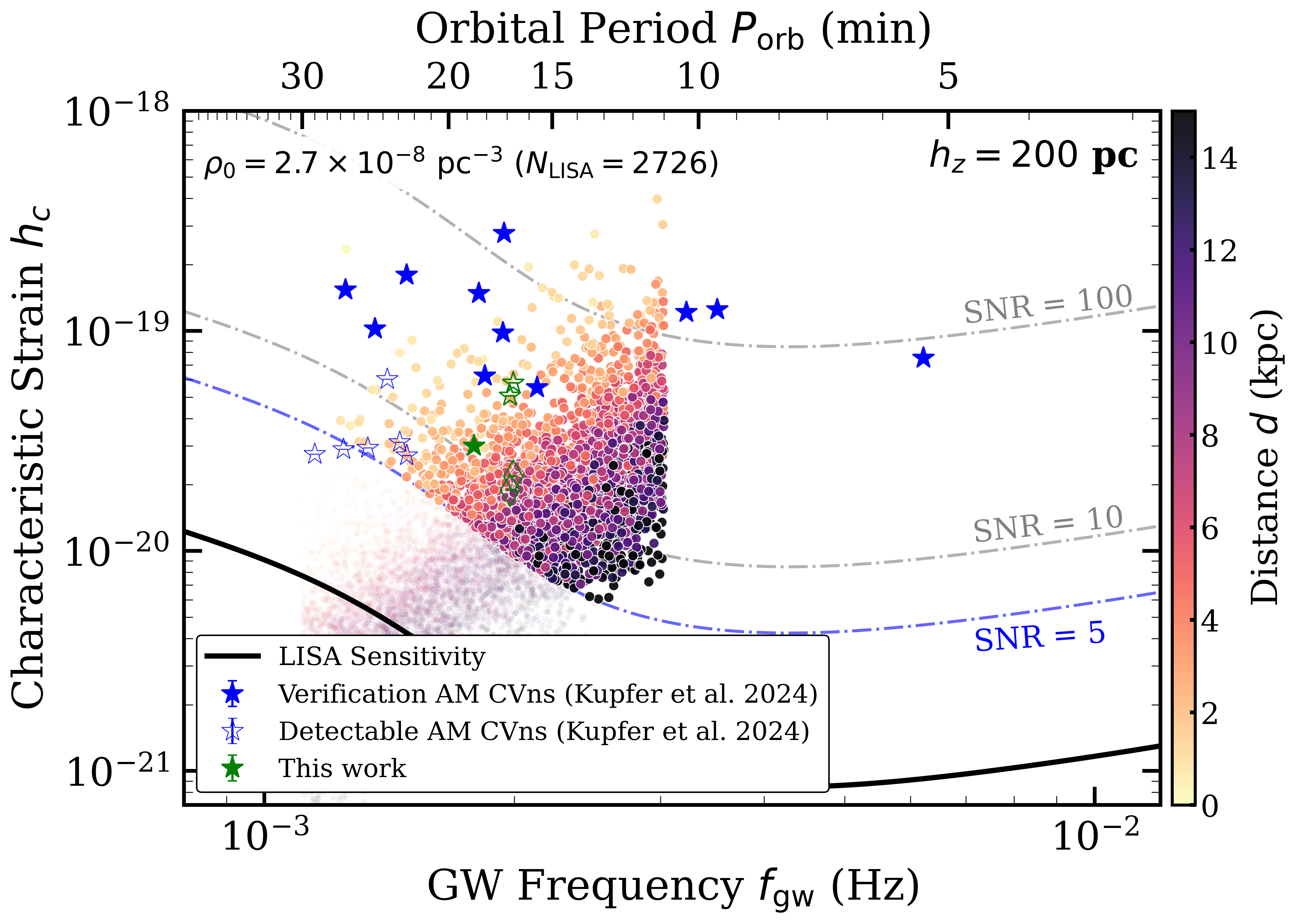} 
    \includegraphics[width=0.49\linewidth]{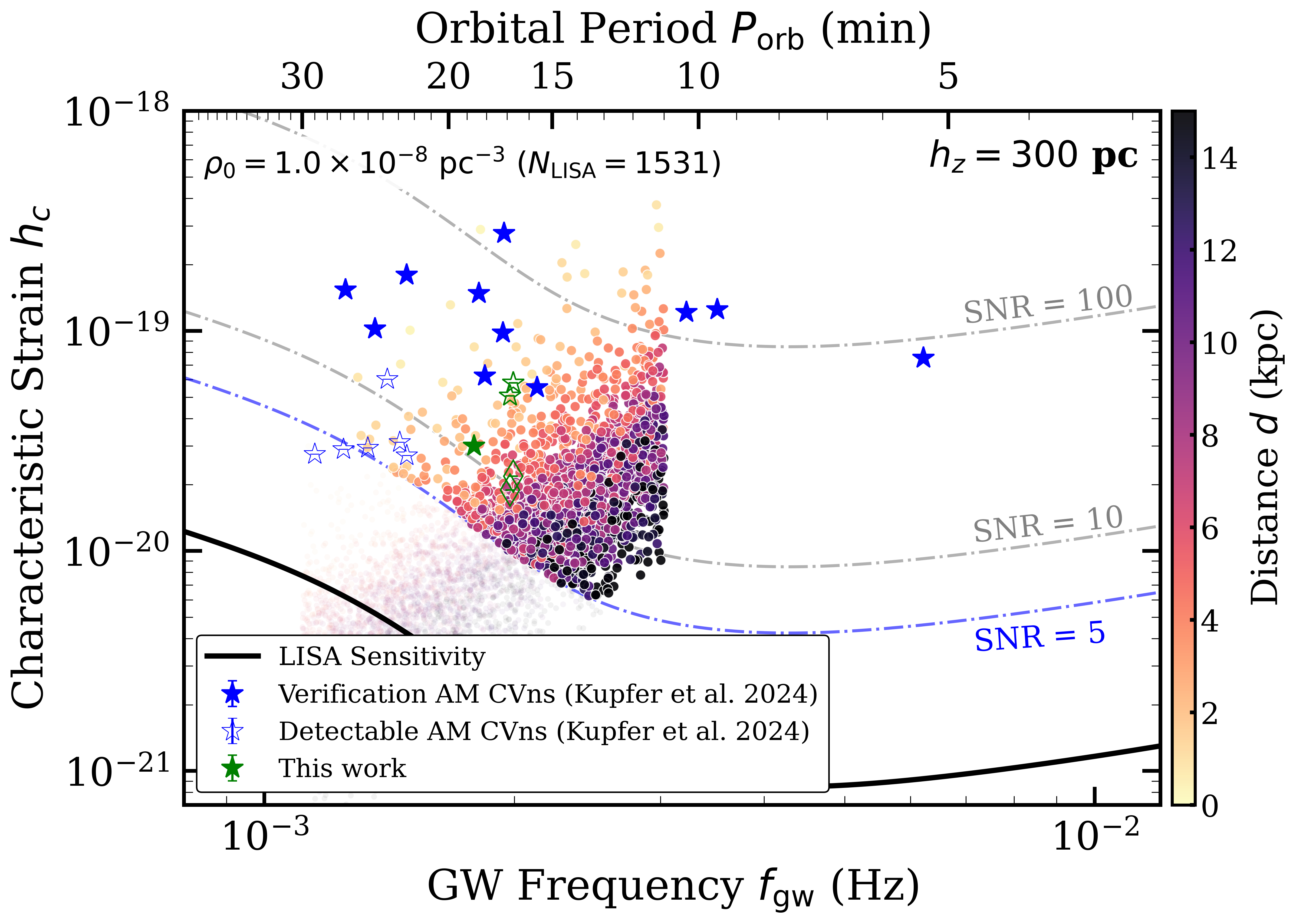}
    \includegraphics[width=0.49\linewidth]{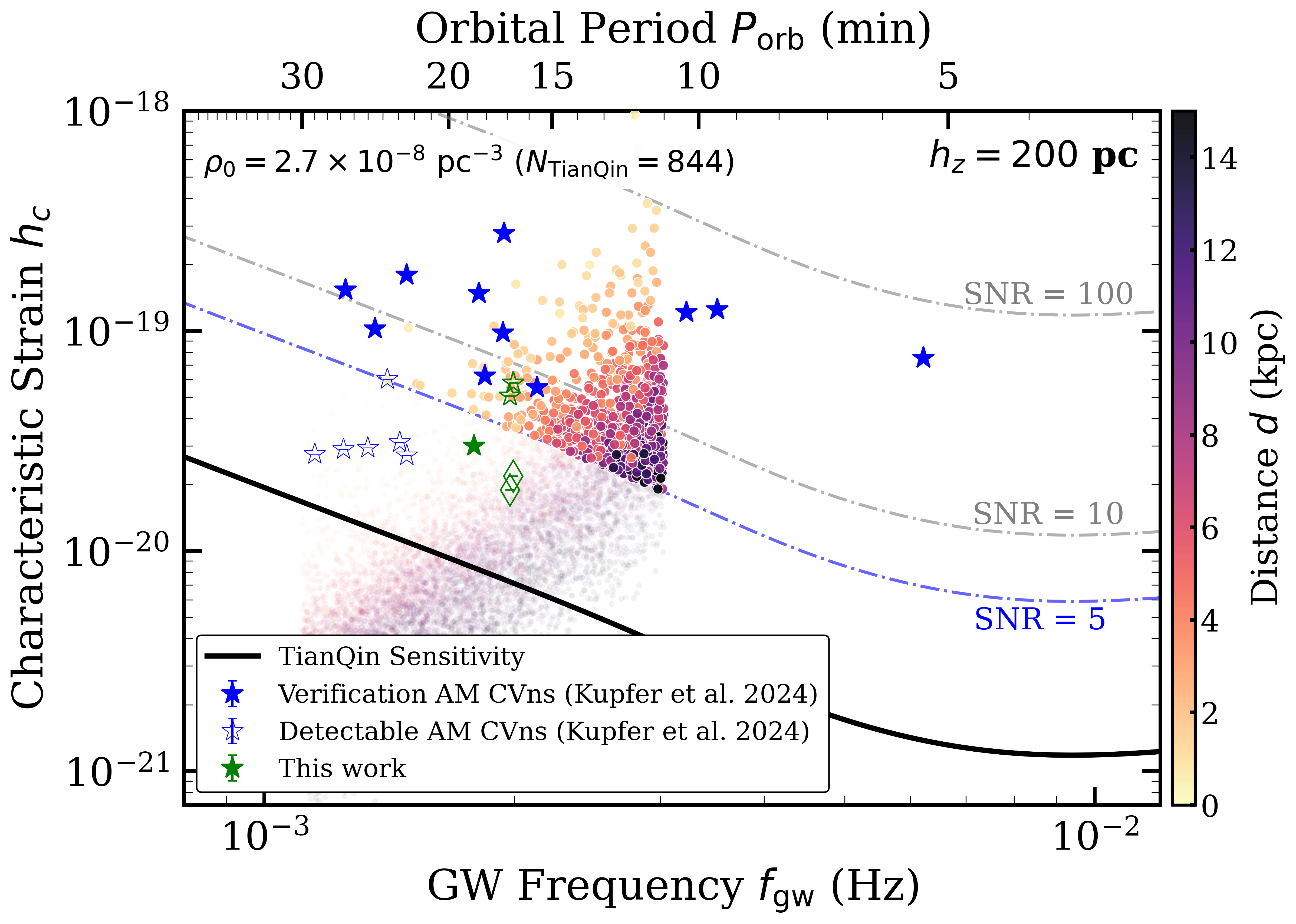} 
    \includegraphics[width=0.49\linewidth]{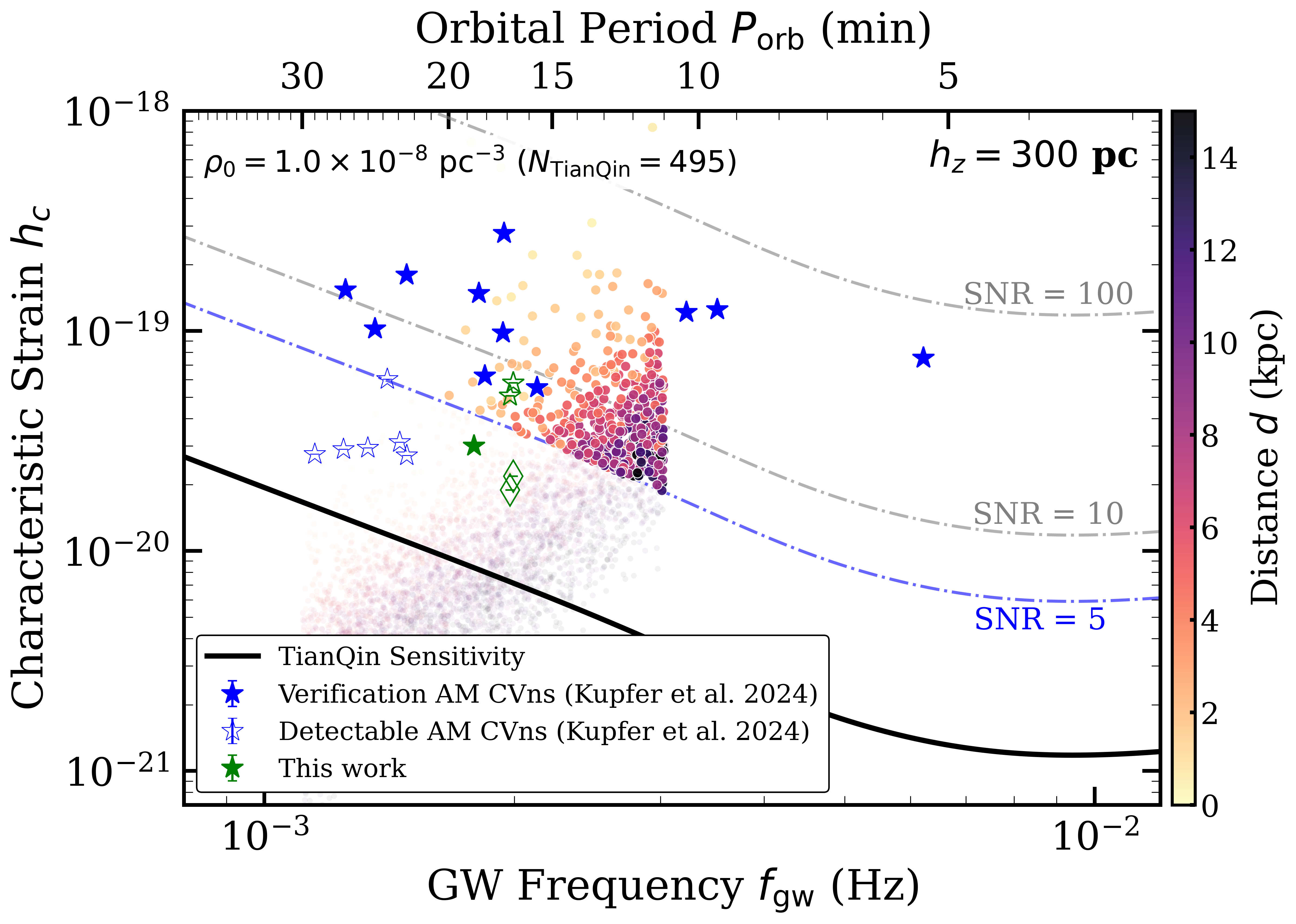}
    \caption{ Same  as  Fig.~\ref{fig:GWpopulation}, but for a helium star donor channel. In this model, about 48\% of the total population is expected to be detectable by LISA and 31\% by TianQin during a 4-year mission  with a signal-to-noise ratio $(S/N) \ge 5$.} 
    \label{fig:GWpopulation_app}
\end{figure*}

\begin{table}[ht]
\centering
\caption{Same as Table ~\ref{tab:amcvn_detections}, but showing results for simulations of the helium star donor channel. } 

\label{tab:amcvn_detections_app}
\renewcommand{\arraystretch}{1.1} 
\setlength{\tabcolsep}{2pt}   
\begin{tabular*}{\columnwidth}{@{\extracolsep{\fill}} c|c|cccc} 
\hline
\hline
Disk scale  & Space density, $\rho_0$ & $N_{\rm total}$ & $N_{\rm LISA}$ & $N_{\rm TianQin}$ & $N_{\rm LSST}$  \\
height, $h_z$ & ($\times 10^{-8}\ \text{pc}^{-3}$) &  &  &  &  \\
\hline

\multirow{3}{*}{200 pc} & 1.2 & 3909 & 1886 & 1215 & 1295 \\
                        & 2.7 & 8617 & 4158 & 2681 & 2893 \\
                        & 4.2 & 13326& 6429 & 4146 & 4629 \\
\hline

\multirow{3}{*}{300 pc} & 0.5 & 2352 & 1134 & 731 & 1029 \\
                        & 1.0 & 4749 & 2290 & 1476 & 1996 \\
                        & 1.5 & 7146 & 3447 & 2223 & 3078 \\
\hline

\hline
\end{tabular*}
\end{table}

The simulations were performed following the approach described in Section~\ref{sec:detectability}. To sample the donor mass, we used the evolutionary track for the helium star channel, where $q \propto P_{\rm orb}^{-1.75}$ (see Fig.~\ref{fig:J2007_fit}, right panel). We assumed that across the $11-30$~minute orbital period range (starting from the period minimum and moving toward longer periods), the mass ratio varies from 0.31 down to 0.06.

Fig.~\ref{fig:GWpopulation_app} shows the characteristic strain versus the GW frequency for a single realization of the high-state AM CVns population for LISA and TianQin, assuming disk scale heights of $h_z=200$~pc and $h_z=300$~pc. The simulated population matches known verification and detectable AM CVn systems more closely than the WD channel simulations at long orbital periods ($11-30$ minutes). This suggests that current electromagnetic surveys may be biased toward detecting the nearest sources and systems evolving via helium star channels.

Table~\ref{tab:amcvn_detections_app} presents the median number of detectable binaries for LISA, TianQin, and LSST. The simulations for the helium star channel suggest that approximately $\approx48\%$ of the total population should be detectable by LISA and $\approx 31\%$ by TianQin during a 4-year mission with a signal-to-noise ratio $(S/N) \ge 5$. LSST is expected to detect $\approx34-44\%$ of the systems during its 10-year survey at a median limit of $m_{r}\lesssim26.9$~mag; this result is similar to the WD channel simulations.

Similarly to the WD channel, we ran additional simulations for the helium star donor channel using the distributions $P_{\rm orb} \propto t^{0.28}$ and $\dot{M} \propto P_{\rm orb}^{-5.68}$ (see Fig.~\ref{fig:models}). Because the orbital period evolution follows a similar trend to the WD channel, we adopted  the same $P_{\rm orb} \propto t^{0.28}$ relation. Based on this helium star model (see Fig.~\ref{fig:models} and Eq.~(\ref{eq:Mdot_crit})), the high-state phase corresponds only to a period range of $11-22$ minutes. Running the simulation with these distributions decreases the number of detectable sources by a factor of $\sim 1.3$ for LISA, $\sim 3.4$ for TianQin, and $\sim 1.1$ for LSST.

\end{appendix}

\end{document}